\documentclass[twoside]{article}

\usepackage[accepted]{aistats2022}
\usepackage[round]{natbib}

\usepackage{graphicx}
\usepackage{amsmath}
\usepackage{natbib, float}
\usepackage{amsfonts}
\usepackage{lscape}
\usepackage{color}
\usepackage{multirow}
\usepackage{multicol}
\usepackage{mathtools}
\usepackage{dsfont}
\usepackage{etoolbox}
\usepackage[caption = false]{subfig} 
\usepackage{enumerate}
\usepackage{comment}
\usepackage[ruled, vlined]{algorithm2e}
\usepackage{todonotes}

\graphicspath{{../figures/}}

\DeclareMathOperator{\pb}{\boldsymbol{p}}

\DeclareMathOperator{\bX}{\boldsymbol{X}}

\DeclareMathOperator{\bbv}{\boldsymbol{v}}
\DeclareMathOperator{\by}{\boldsymbol{y}}

\DeclareMathOperator{\bx}{\boldsymbol{x}}

\DeclareMathOperator{\x}{\mathbf{x}}

\DeclareMathOperator{\btheta}{\boldsymbol{\theta}}

\DeclareMathOperator{\Ex}{\mathbb{E}}

\DeclareMathOperator{\0}{\boldsymbol{0}}

\newcommand{\norm}[1]{\left\lVert#1\right\rVert} 

\begin{document}

%

%

\twocolumn[

\aistatstitle{Lagrangian Manifold Monte Carlo on Monge Patches}

\aistatsauthor{ Marcelo Hartmann \And Mark Girolami \And Arto Klami }

\aistatsaddress{University of Helsinki \\ Department of Computer Science \And  University of Cambridge \\ Department of Engineering \\ \& The Alan Turing Institute \And University of Helsinki \\ Department of Computer Science} ]

\begin{abstract}
The efficiency of Markov Chain Monte Carlo (MCMC) depends on how the underlying geometry of the problem is taken into account. For distributions with strongly varying curvature, Riemannian metrics help in efficient exploration of the target distribution. Unfortunately, they have significant computational overhead due to e.g. repeated inversion of the metric tensor, and current geometric MCMC methods using the Fisher information matrix to induce the manifold are in practice slow. We propose a new alternative Riemannian metric for MCMC, by embedding the target distribution into a higher-dimensional Euclidean space as a Monge patch and using the induced metric determined by direct geometric reasoning. Our metric only requires first-order gradient information and has fast inverse and determinants, and allows reducing the  computational complexity of individual iterations from cubic to quadratic in the problem dimensionality. We demonstrate how Lagrangian Monte Carlo in this metric efficiently explores the target distributions. 
\end{abstract}

\section{INTRODUCTION}

Markov Chain Monte Carlo (MCMC) algorithms provide samples from complex distributions for which direct sampling is difficult, and are routinely used in Bayesian statistics for sampling from the posterior distribution of a model \citep{chkrebtii:2016, calder12}. The conditions for asymptotically valid samplers are mild, but
efficiently exploring high-dimensional distributions remains a major challenge. Modern methods typically convert the problem into numerical integration of an augmented dynamic system, based e.g. on Langevin diffusion
 \citep{roberts:1996, robert:2002, green:2015}, Hamiltonian Dynamics \citep{duane:1987, neal:2011, betan:2017} or Lagrangian dynamics \citep{fang:2014, lan:2015}.

The augmented dynamics combine the logarithm of the target distribution with a kinetic term and simulate the time-evolution of the system. By using gradient information to drive the evolution they both convergence to the target distribution faster and improve exploration of the likely set.
However, high-dimensional problems with strong correlations between individual dimensions and/or vastly different marginal variances are still challenging \citep{robert:2002,betan:2017}. To an extent this can be addressed by tuning a mass matrix $M$ controlling the kinetic energy to globally de-correlate the parameter's dependency. This is equivalent to changing the \emph{metric} on the parameter space, but still assuming some Euclidean metric \citep{neal:2011}. However, every global metric is necessarily a compromise between efficiency in regions of low curvature and accurate exploration of regions of high curvature.

Geometric MCMC algorithms \citep{girolami:2011, xifara:2014, lan:2015, betan:2017, beskos:2017} use differential geometry to account for local curvature,
replacing the mass matrix $M$ with position-dependent matrix $G(\bx)$ that is the metric tensor of a suitable Riemannian manifold. Accounting for the local curvature improves the efficiency of the sampler especially in high-curvature regions 
\citep[see][for many examples]{xifara:2014, girolami:2011, beskos:2017}. 
The choice of the manifold and hence the metric is free, but existing literature focuses almost solely
on the manifold and metric induced by the Fisher Information (FI) matrix of an underlying probabilistic model \citep{mark:2011}. It is a natural choice that can be derived from local Kullback-Leibler divergence, but only applicable for the specific case of posterior sampling
as it is derived from a probabilistic model which mimics
random variation in real data-sets.

The improved exploration comes with significant computational cost, and hence geometric MCMC methods are not widely used in practice. As the metric tensor is position-dependent, we now need to compute and invert it in every step of the numerical integration, sometimes several times.
Already forming the FI matrix is demanding as it requires expected second derivatives of the log density of the model, and inversion has cubic complexity in the problem dimensionality $D$.

We present a new Riemannian metric that also relates to local curvature of the distribution but that is computationally efficient and generally applicable, based on pure geometric reasoning rather than relying on statistical properties of a model. We propose an embedding based on the graph of the target distribution $\pi_{\bX}$ as a manifold in a higher-dimensional Euclidean space, using a scaled Monge parameterization $\Xi(\bx) = (\bx, \alpha \log \pi_{\bX}(\bx))$. The manifold is generated by the \emph{Monge patch} embedding named after Gaspard Monge, one of the inventors of differential geometry \citep{oneill:2006}.
This operation defines a Riemannian manifold
with a natural metric tensor. The metric tensor $G_M(\bx)$ is expressed as rank-one perturbation of the identity matrix with the rank-one term being the outer product of the gradients of the log target density. Consequently, it has efficient closed-form inverse
as well as efficient closed-form determinant, offering significant computational savings.

The new metric captures the local curvature of the target density directly via simultaneous relations between the second fundamental form of the manifold, the Hessian of the target density and the Christoffel symbols. It provides similar advantages in exploration of complex regions of the distribution as the Fisher metric, and in expectation can be interpreted as regularized FI matrix. The control parameter $\alpha$ allows fine-tuning the embedding and the metric for overall computational efficiency.

The metric is general and applicable for various geometric MCMC algorithms. We demonstrate it with the Lagrangian Monte Carlo (LMC) \citep{lan:2015}. Compared to Riemannian manifold HMC (RMHMC), LMC has the advantage of an explicit numerical integrator that only requires two matrix inversions per iteration. However, it is not symplectic (volume-preserving) and hence requires also computing determinant adjustment for the proposals acceptance check. The costly computation of the determinants and \emph{Christoffel symbols} required for the numerical integrator have limited the interest in LMC, but in our metric both can be computed efficiently.
In our experiments, LMC in the Monge metric outperforms algorithms operating in Euclidean or Fisher metrics.

\section{BACKGROUND}\label{sec:basics}

We briefly summarize Hamiltonian Monte Carlo (HMC) as an example algorithm using augmented dynamics and discuss the role of metrics for the simulation. We then provide the foundations of differentiable manifolds, introducing concepts relating to curvatures of the manifolds and their relationship to metrics.
\subsection{Hamiltonian Monte Carlo and Metrics}

Hamiltonian Monte Carlo \citep{neal:2011} provides samples from a probability distribution $\pi_{\bX}(\bx)$ by simulating the time-evolution of the Hamiltonian
\[
H(\bx,\pb) = -\log \pi_{\bX}(\bx) + \frac{1}{2} \log |M| + \frac{1}{2} \pb^T M^{-1} \pb
\]
where the momentum variables $\pb$ are sampled (typically) from a normal distribution. A new proposal is generated by simulating the trajectory of the pair $(\bx, \pb)$ using numeric integration that alternates between updates for the position $\bx$ and the momentum $\pb$. This simulation is done for $L$ iterations before determining whether the proposal is accepted. Variants of HMC, such as the No-U-Turn-Sampler \citep[NUTS;][]{hoffman:14}, are today the most common methods for statistical inference
and are widely implemented in probabilistic programming languages.

The efficiency of HMC depends on the choice of the \emph{mass matrix} or \emph{metric tensor} $M$, which is typically tuned during warm-up. For instance, $M$ proportional to the covariance of the target distribution effectively de-correlates the dimensions and improves exploration \citep{neal:2011}. However, no global metric can help coping with differences in local stretching or squeezing of the manifold, and hence techniques like explicit reparameterization are used for complex distributions \citep{Papaspiliopoulos:2007}.

Rather than using a global metric, we can conduct HMC on Riemannian manifolds (RMHMC) by using a position-dependent metric tensor $G(\bx)$ instead \citep{girolami:2011}. This allows coping with changes in local curvature, assuming the metric is chosen suitably. This extension results in an implicit numerical integrator since two of the updates have the same variable on both sides:
\begin{align*}
\pb^{(n+1/2)} &= \pb^{(n)} - \frac{\epsilon}{2} \nabla_{\bx} H \left (\bx^{(n)}, \pb^{(n+1/2)}
\right ),\\
\bx^{(n+1)} &= \bx^{(n)} + \frac{\epsilon}{2} 
\left [
\nabla_{\pb} H \left (\bx^{(n)}, \pb^{(n+1/2)}
\right ) \right . \\
&\quad\quad\quad\quad\quad + \left . \nabla_{\pb} H \left (\bx^{(n+1)}, \pb^{(n+1/2)}
\right )
\right ],
\\
\pb^{(n+1)} &= \pb^{(n+1/2)} - \frac{\epsilon}{2} 
\nabla_{\bx} H \left (\bx^{(n+1)}, \pb^{(n+1/2)}
\right ).
\end{align*}
The solution of these equations requires matrix inversion during every iteration since $\nabla_{\bx} H(\bx,\pb) = G(\bx)^{-1} \pb$. Furthermore, the implicit equations are solved by a fixed-point iteration and hence there is a need of computing inverse matrices multiple times.
Usually the metric is derived from FI, as explained in more detail in Section~\ref{sec:FI}. MCMC chains in Fisher metric behave better compared to any Euclidean metric, but the extensive computational cost and difficulty of computing the metric has prevented wide-spread use of RMHMC. \citet{paquet:2018} considered using the Hessian of the target density as the metric tensor as an alternative, but it has the same computational cost.

In Section~\ref{sec:LMCe} we will consider in detail a variant of RMHMC, Lagrangian Monte Carlo \citep{lan:2015}, that avoids implicit equations but requires calculation of determinants and Christoffel symbols instead.

\subsection{Differential Geometry Preliminaries} \label{sec:diffgeom}


Our point of departure is the notion of a {\it differentiable manifold}.
We call a set $\mathcal{M}$ a {\it differentiable manifold} of dimension $m$ (in short manifold) if together with bijective mappings (also called parametrizations or system of coordinates) $\Xi_i (x_1, \ldots, x_m) : \mathcal{X}_i \subset \mathbb{R}^m \rightarrow \mathcal{M}$ where $\mathcal{X}_i$ is a chart, they satisfy,
\begin{enumerate}
\item[$(a)$] $\bigcup_{i} \Xi_i(\mathcal{X}_i) = \mathcal{M}$
\item[$(b)$]  For each $i$, $j$, $\Xi_i(\mathcal{X}_i) \bigcap \Xi_j(\mathcal{X}_j) \neq \emptyset$ and that $\Xi_i^{-1} \circ \Xi_j$ are differentiable mappings.
\end{enumerate}
The family $(\Xi_i, \mathcal{X}_i)$ is also called a differentiable structure on $\mathcal{M}$, and allow us to extend notions of the differential calculus in Euclidean space to more general spaces such as some abstract set $\mathcal{M}$ (e.g. a family of probability distributions). 

One of the aims of differential geometry is to enable characterizing the {\it rate of change} for computing derivatives on $\mathcal{M}$ intrinsically, without referring to any external coordinate space. For this we need the notion of a {\it tangent space}.
To do so, consider two overlapping curves that trace out different paths on $\mathcal{M}$ but intersect in a unique point $p \in \mathcal{M}$. With the aid of two distinct charts for each path, we denote $\gamma_1 := \Xi(t) : I_1 \subset \mathbb{R} \rightarrow \mathcal{M}$ and $\gamma_2 := \Pi(t) : I_2 \subset \mathbb{R} \rightarrow \mathcal{M}$. By taking the usual derivatives w.r.t to the variable $t$ at $t_1$ such that $\gamma_1(t_1) =p$ and for the second curve at $t_2$ such that $\gamma_2(t_2) = p$, we obtain 
\begin{align*}
\dot{\gamma}_1 = \sum_{k = 1}^n \frac{\mathrm{d} x_k}{\mathrm{d} t} \frac{\partial}{\partial x_k} \Xi \ \ \ \text{and} \ \ \dot{\gamma}_2 = \sum_{k = 1}^n \frac{\mathrm{d} y_k}{\mathrm{d}t}  \frac{\partial}{\partial y_k} \Pi.
\end{align*}
Because of condition $(b)$ in the manifold definition, the set of vectors $\{\partial/\partial_k \ \Xi \}_k$ and $\{ \partial/\partial_k \ \Pi \}_k$ span the same linear subspace of $\mathbb{R}^n$ at $p \in \mathcal{M}$, differing only in the basis vectors.
Henceforth, we call this linear subspace as {\it tangent space} at $p$, in short $T_p \mathcal{M}$. To see this more clearly, define a new chart $\psi = \Xi \circ h  : I_3 \rightarrow \mathcal{M} $  where $ h =  \Xi^{-1} \circ \Pi$ and note that
\begin{align*}
\frac{\partial}{\partial y_k} \psi = \frac{\partial}{\partial y_k} \Xi \circ h = \sum_{c = 1}^n  \frac{\mathrm{d} x_c}{\mathrm{d} y_k} \frac{\partial}{\partial x_c} \Xi
\end{align*}
for $k = 1, \ldots, n$. 

Since the Jacobian of transformation $h$ does not vanish for any $p \in \mathcal{M}$, we have $\{ \partial /\partial y_k \ \psi \}_k$ and $\{ \partial/\partial x_k \ \Xi \}_k$ as the only different basis vectors of the set $T_p\mathcal{M}$. Furthermore, we can define an inner product of elements of the space $T_p \mathcal{M}$ as $g : T_p\mathcal{M} \times T_p\mathcal{M} \rightarrow \mathbb{R}$ and then note that $g$ is invariant with respect of different charts of the manifold.
\citet{gauss:1902} noted the implications of this alreay in 1827:
If we want to study the curvature (how much $\mathcal{M}$ deviates from a Euclidean space, or how it stretches and squeezes locally) of the set $\mathcal{M}$, it is enough to know the metric $g$ -- we do not need the exact form of the charts.

\begin{figure}
\begin{center}
\includegraphics[width=0.7\columnwidth]{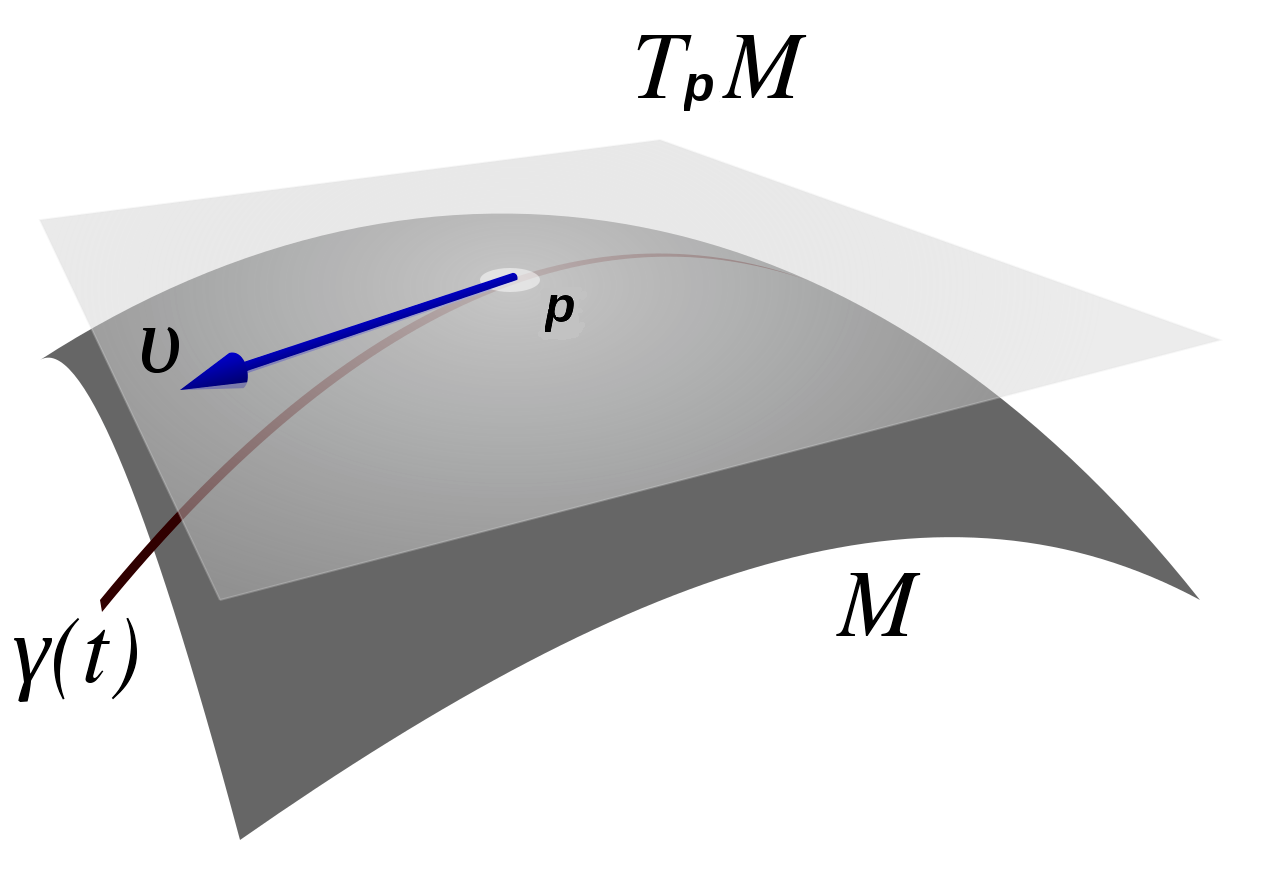}
\caption{Manifold $\mathcal{M}$ and its tangent space $T_{p}\mathcal{M}$, with $v$ being a tangent vector of the curve $\gamma(t)$.}
\end{center}
\end{figure}

\subsection{Riemannian Manifolds and Metrics}
\label{sec:FI}

A {\it Riemannian manifold} is a manifold which associates for each point $p  \in \mathcal{M}$ an inner product $g$ (symmetric, bilinear and positive-definite) for the vectors in $T_p \mathcal{M}$. For a given parametrization $\Xi$ and tangents $v = s_i \ \partial/\partial_i \ \Xi$ and $u = t_i \ \partial/\partial_i \  \Xi$, we have $g(u, v) = \left\langle u, v \right\rangle_p = \boldsymbol{s}^\top G(p) \boldsymbol{t}$ where the coefficients of the metric $g$ are the
 elements of the positive-definite matrix $G(p)$ given by the inner products
\begin{align*}
G_{i, j}(p) = \left\langle  \frac{\partial}{\partial x_i} \Xi, \frac{\partial}{\partial x_j} \Xi \right\rangle_p  \ \ \mathrm{and} \ \ \boldsymbol{s}, \boldsymbol{t} \in \mathbb{R}^n.
\end{align*}
%
Such a matrix is called {\it metric tensor}.
In this way Riemannian manifolds can be directly defined by a differentiable structure on a set $\mathcal{M}$ and a positive-definite matrix $G$ at each $p \in \mathcal{M}$,
without reference to any specific system of coordinates. 

One particular Riemannian metric used broadly in statistics and machine learning uses the Fisher information matrix as the metric tensor \citep{amari:2019,girolami:2011,lan:2015}. In context of MCMC, it provides a metric that accounts for a probabilistic model for data, but that requires computing the expectation of the Hessian that is often difficult \citep{yudi:2001}. If the probabilistic model satisfies suitable regularity conditions \citep{mark:2011}, we can express the metric as
\begin{align*}
G_{i, j}(p) &= \Ex_Y\left(\frac{\partial}{\partial p_i} \log \pi_Y(Y|p) \frac{\partial}{\partial p_j} \log \pi_Y(Y|p) \right) \nonumber \\
&= - \Ex_Y\left( \frac{\partial^2}{\partial p_i \partial p_j} \log \pi_Y(Y|p) \right)  \nonumber \\
& = - \int_\Omega \frac{\partial^2}{\partial p_i \partial p_j} \log \pi_Y(y|p) \pi_Y(y|p) \hspace{0.03cm} \mathrm{d}y.
\end{align*}
where $Y$ is a random variable (data yet to be observed), $y$ is the observed data and $\Omega$ is the space of all possible data outcomes. We call the resulting metric \emph{Fisher metric} and denote the metric tensor by $G_F(\cdot)$.

FI  characterizes the lower bound of the variance of unbiased estimators and it can also be derived from the Kullback-Leibler divergence between two probability distribution of the same family and hence offers interesting theoretical connections, but ultimately the choice has still been primarily justified by good empirical properties \citep{girolami:2011,betan:2017}. Finally, it is only applicable for posterior sampling and not for general sampling problems.



\section{MONGE PATCH AND METRIC}

Our goal is to form a metric that accounts for local curvature of the target distribution, but is (a) computationally efficient and (b) applicable for general target densities, rather than requiring an underlying probabilistic model for forming the metric. We seek for such a metric based on pure geometric principles of hyper-surfaces embedded in higher-dimensional Euclidean spaces \citep{gauss:1902,docarmo:1992,docarmo:2016}.



Let $\mathcal{M}$ and $\mathcal{N}$ be manifolds of dimension $m$ and $n$ respectively with $m \leq n$. We say $\mathcal{M}$ is an {\it embedding} if for a differentiable mapping $\varphi : \mathcal{M} \rightarrow \mathcal{N}$ the differential $\mathrm{d}\varphi_p(v) : T_p\mathcal{M} \rightarrow T_{\varphi(p)}\mathcal{N}$ is injective and $\varphi$ is a bijection. 
Consider a target probabilistic model $\bX \sim \pi_{\bX}(\cdot)$ from which we would like to obtain samples from and denote its logarithm as $\ell(\bx) = \log \pi_{\bX}(\bx) : \mathcal{X}\subseteq \mathbb{R}^D \rightarrow \mathbb{R}$. 

We then represent the manifold $\mathcal{M}$ as an embedding using the target distribution to define the embedding in $\mathcal{N}$ (which is a subspace of the $D+1$ dimensional Euclidean space) with $\varphi$ as the identity function. The embedding is,
\begin{equation*}
\Xi(\bx) = (\bx, \alpha \ell(\bx)) \in \mathcal{M}
\label{eq:embedding}
\end{equation*}
and thus $\mathcal{M} =\{ (z_1, \ldots, z_{D+1}) =: \Xi(\bx) \in (\mathcal{X} \times \mathbb{R}) \subset \mathbb{R}^{D + 1} : \bx \in \mathcal{X} \subset \mathbb{R}^D \}$ is the embedded manifold via the \emph{scaled Monge patch} $\Xi$ with $\alpha \ge 0$. This extends the Monge parameterization $(\bx, \ell(\bx))$ with a parameter $\alpha$ that will be used for controlling the curvature information of the induced metric. Alternatively, we can interpret this as embedding of the logarithm of the tempered distribution $\pi_{\bX}(\bx)^{\alpha}$. This embedding is arbitrary in the sense that we have no specific rationale for the choice, but as will be shown next it induces a metric that has several desirable properties. 


For a tangent $v = \sum_{i = 1}^n s_i \ \partial/\partial x_i \ \Xi(\bx) \in T_p\mathcal{M}$ where $$\dfrac{\partial}{\partial x_i} \Xi(\bx) = \bigg( \underbrace{0, \ldots, 1, \ldots, 0}_{i^{th} \ \mathrm{position}}, \alpha \frac{\partial}{\partial x_i} \ell(\bx) \bigg),$$ we obtain that $\mathrm{d} \varphi_p (v)$ is injective $\forall p$. Therefore, as defined previously for tangents $u, v \in T_p \mathcal{M}$ the metric induce by this embedding becomes,
\begin{align*}
& \hspace{0.7cm} g_M(u, v) = \boldsymbol{s}^\top G_M(p) \boldsymbol{t} \\[0.2cm]
&= \boldsymbol{s}^\top \begin{bmatrix}
\displaystyle\sum_{d = 1}^{D} \dfrac{\partial \hspace{0.02cm} \Xi_d}{\partial {x_1}} \dfrac{\partial \hspace{0.02cm} \Xi_d}{\partial {x_1}} & \cdots & \displaystyle\sum_{d = 1}^{D} \dfrac{\partial \hspace{0.02cm} \Xi_d}{\partial {x_1}} \dfrac{\partial \hspace{0.02cm} \Xi_d}{\partial {x_D}} \\ 
\vdots & \ddots & \vdots \\ 
\displaystyle\sum_{d = 1}^{D} \dfrac{\partial \hspace{0.02cm} \Xi_d}{\partial {x_D}} \dfrac{\partial \hspace{0.02cm} \Xi_d}{\partial {x_1}} & \cdots & \displaystyle\sum_{d = 1}^{D} \dfrac{\partial \hspace{0.02cm} \Xi_d}{\partial {x_D}} \dfrac{\partial \hspace{0.02cm} \Xi_d}{\partial {x_D}}
\end{bmatrix}   \boldsymbol{t} \\[0.2cm]
&= \boldsymbol{s}^\top \begin{bmatrix}
1 + \alpha^2 \frac{\partial}{\partial x_1} \ell(\bx)^2\ & \cdots & \alpha^2 \frac{\partial}{\partial x_1} \ell(\bx) \frac{\partial}{\partial x_n} \ell(\bx) \\ 
\vdots & \ddots & \vdots \\ 
\alpha^2 \frac{\partial}{\partial x_n} \ell(\bx) \frac{\partial}{\partial x_1} \ell(\bx) & \cdots & 1 + \alpha^2 \frac{\partial}{\partial x_n} \ell(\bx)^2
\end{bmatrix}   \boldsymbol{t} 
\end{align*}
where each $\bx$ in the support of our target (or the domain of the target distribution) uniquely determines a specific point $p$ on the manifold $\mathcal{M}$. Hence we denote the metric tensor in matrix form as
%
\begin{align}\label{eq:metrics}
G_M(\bx) &=  I_D + \alpha^2 \nabla \ell(\bx) \nabla \ell(\bx)^\top,  
\end{align}
slightly abusing the notation to express it directly in terms of $\bx$.
The matrix \eqref{eq:metrics} is symmetric and positive-definite and hence the pair $(\mathcal{M}, g_M)$ is a Riemannian manifold. We call the resulting metric the \emph{Monge metric}.

\subsection{Interpretation}

The local geometric properties of 
manifolds
have two important quantities with direct interpretation: 
the \emph{first fundamental form} 
which relates to lengths of the curves on $\mathcal{M}$ and the \emph{second fundamental form} that relates to the curvature, i.e., how much the manifold locally deviates from the Euclidean space (or the tangent plane).
Both the Fisher metric and the Monge metric are connected to the first fundamental form as they tell us a way to measure lengths of curves on $\mathcal{M}$. In some statistics literature the Fisher metric has been linked with the idea of curvature, see \citet{calder12}, \cite{girolami:2011} and \cite{paquet:2018}, due to the its definition as the expected value of the Hessian matrix. However, the Monge metric has natural geometric reasoning as it additionally has a direct notion of curvature due to the clear manifestation of Hessian matrix of $\ell$ in the second fundamental form on the embedded Riemannian manifold $\mathcal{M}$.



The second fundamental form $g_* = \left\langle \ddot\gamma_1(\bx(t_1)), N(\bx) \right\rangle$ is formally defined as the inner product between the acceleration of a curve on the manifold, $\ddot\gamma_1(\bx(t_1))$, and the normal vector 
\begin{align*}
N(\bx) =  -\frac{\big( \alpha \frac{\partial}{\partial x_1} \ell(\bx), \ldots, \alpha \frac{\partial}{\partial x_D} \ell(\bx), 1\big)}{\sqrt{1 + \alpha^2 ||\nabla \ell(\bx)||^2}},
\end{align*}
for every $p \in \mathcal{M}$.
%
After algebraic manipulation
and cancelling the terms of 
$\dot\gamma_1(\bx(t_1))$ orthogonal to $N(\bx(t_1))$ and dropping the notation of the argument $t_1$, we obtain \citep[see][]{press:2010, docarmo:2016}
\begin{align*}
g_* =
\boldsymbol{s}^\top \left\lbrace \left\langle\frac{\partial^2}{\partial x_i \partial x_j}\Xi(\bx), N(\bx) \right\rangle \right\rbrace_{i, j} \boldsymbol{s}
\end{align*}
Thus we can see that
\begin{align*}
g_*  &= \boldsymbol{s}^\top \frac{\alpha }{c}
\underbrace{\begin{bmatrix}
\frac{\partial^2}{\partial x_1 \partial x_2} \ell(\bx) & \cdots & \frac{\partial^2}{\partial x_1 \partial x_D} \ell(\bx) \\ 
\vdots & \ddots & \vdots \\ 
\frac{\partial^2}{\partial x_D \partial x_1} \ell(\bx) & \cdots & \frac{\partial^2}{\partial x_D \partial x_D} \ell(\bx)
\end{bmatrix}}_{H(\bx)} \boldsymbol{s},
\end{align*}
where $c=\sqrt{1 + \alpha^2 ||\nabla \ell(\bx)||^2}$ and $H(\bx) = \nabla^2\ell(\bx)$ is the Hessian matrix of the logarithm of the target distribution. 
The curvature of the Monge metric, as measured by the second fundamental form, is hence a scaled version of the Hessian that encodes local scaling and stretching information. This provides an intuitive and natural interpretation for the metric, even though it was induced by a seemingly arbitrary embedding.

The Monge metric is derived from a different perspective than the Fisher metric, but they are related. For the case where
the logarithm of the target distribution is $\ell(\bx) = \log \pi_Y (Y|\bx)$ and $\pi_Y(\cdot|\bx)$ is a model that defines the random generating mechanism of the data, we obtain
$
\alpha^{-2}
\Ex_Y\left(G_M(\bx)\right) = \alpha^{-2} I_D + G_F(\bx)
$
by computing the expectation of the negative Hessian 
over $Y$. 
That is, in expectation the metric can be seen as biased or regularized estimator for FI, so that inverse $\alpha$ controls the regularization.

\begin{figure}
    \centering
    \includegraphics[width=\columnwidth]{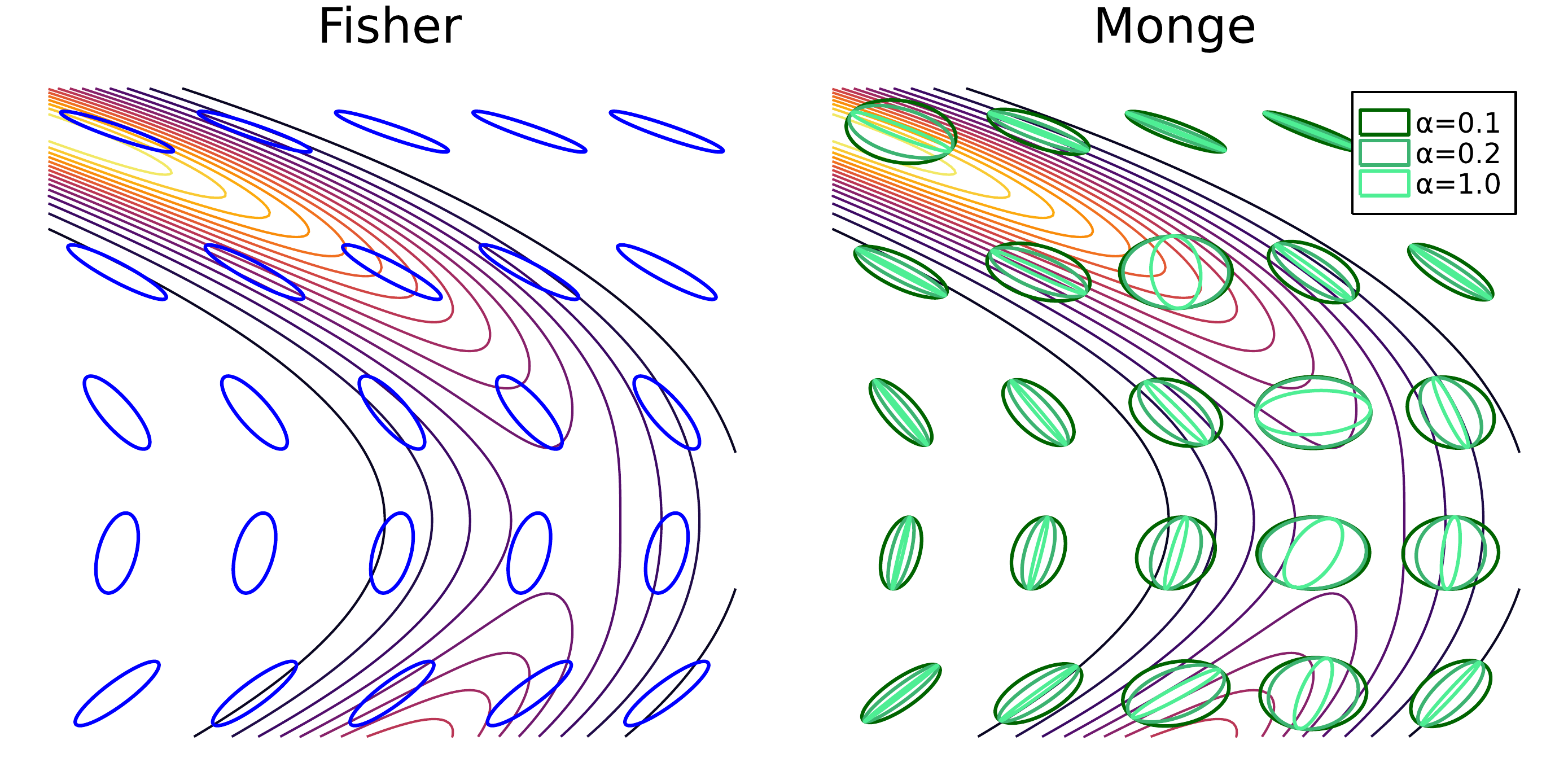}
    \caption{Illustration of the metric on the Rosenbrock distribution. The Monge metric captures the shape of the distribution in similar manner as the Fisher metric. The $\alpha$ parameter controls the  embedding and scales the metric; with small $\alpha$ (dark green) it is close to Euclidean and with large $\alpha$ (light green) we get very elongated metric for areas of high curvature.}
    \label{fig:illustration}
\end{figure}

Figure \ref{fig:illustration} illustrates 
 the Monge and Fisher metrics for a banana-shaped posterior ($\pi_{\bX}(x_1, x_2) \propto \prod_i \mathcal{N}(y_i|x_1 + x_2^2, \sigma^2_y) \mathcal{N}(x_1|0, \sigma^2)\mathcal{N}(x_2|0, \sigma^2)$  with $\sigma_y^2 = \sigma^2 = 0.5$ and $n = 10$ observations $y_i$).
Here the Fisher metric is constant w.r.t. to the $x_1$ coordinate \citep{bornn:2011}, whereas the Monge metric is bivariate. The Monge metric becomes identity at the mode, and flattens towards spherical Euclidean metric for $\alpha \rightarrow 0$. Outside the mode it behaves similarly to the Fisher metric, but for large $\alpha$ is more elongated.
The question of optimal $\alpha$ is an empirical one.

\subsection{Computation}

\paragraph{Fast Inverse and Determinants}
The metric \eqref{eq:metrics} has efficient inverse via Sherman-Morrison lemma as
\[
G_M(\bx)^{-1} = I_D - \alpha^2 \frac{\nabla \ell(\bx) \nabla \ell(\bx)^\top}{1 + \alpha^2 \norm{\nabla \ell(\bx)}^2}
\]
with $\mathcal{O}(D^2)$ complexity. Similarly, the determinant is
\[
\det G_M(\bx) = 1 + \alpha^2 \norm{\nabla \ell(\bx)}^2
\]
with linear complexity. Both are significant improvements over $\mathcal{O}(D^3)$ for general operations in the original LMC formulation \citep{lan:2015}.

\paragraph{Fast Christoffel symbols} The \emph{Christoffel symbols} $\Gamma^k_{i, j}(\bx)$ measure the magnitude of the basis vector
$\partial/\partial x_k \hspace{0.03cm} \Xi$ in the rate of change of the vector $\partial/\partial x_j \hspace{0.03cm} \Xi$ at the direction of $\partial/\partial x_i \hspace{0.03cm} \Xi$ for every point $\Xi(\bx) = p$ of the manifold. Formally the Christoffel symbols are defined as the coefficients of the \emph{Levi-Civita connection} of the Riemannian manifold.

They are required for some algorithms operating on Riemannian manifolds and, if not obtained in closed-form, their computation might incur a significant computational cost in general case. Since $G_M(\bx)$ is obtained using an embedding $\Xi$, we can re-write the Christoffel symbols following \cite[][page 56, equation (10)]{docarmo:1992} as
%
\begin{align*}
\Gamma^k_{i, j}(\bx) &=
\sum_{l = 1}^D G^{-1}_{k, l} (\bx) \left\langle \frac{\partial^2}{\partial x_i \partial x_j}  \hspace{0.03cm} \Xi, \frac{\partial}{\partial x_l} \hspace{0.03cm} \Xi \right\rangle \nonumber \\
& = \frac{\alpha^2}{1 + \alpha^2 \norm{\nabla \ell(\bx)}^2} \frac{\partial}{\partial x_k} \ell (\bx) \frac{\partial^2}{\partial x_i \partial x_j}  \ell (\bx)
\end{align*}
%
using the elements $\tfrac{\partial^2}{\partial x_i \partial x_j} \ell (\bx)$ of the second fundamental form. Even though the metric tensor only requires the gradients, we see that second-order derivatives are needed for computing the Christoffel symbols.

\section{LMC ON MONGE PATCHES} \label{sec:LMCe}
 
The Monge metric is general and applicable for several geometric MCMC methods. We demonstrate it here
for Lagrangian Monte Carlo \citep{lan:2015}, providing detailed derivations in the Supplement.

\paragraph{Lagrangian Monte Carlo}
As explained in Section~\ref{sec:basics}, RMHMC involves two implicit equations that require fixed-point iterations and hence multiple inversions of the metric tensor during every update. An explicit integrator can be developed by switching to \emph{Lagrangian dynamics} and working with \emph{velocity} $\bbv = G(\bx)^{-1}\pb$ instead of the momentum, resulting in Riemannian manifold Lagrangian Monte Carlo \citep{lan:2015}. The energy functional and dynamics are
\begin{align}
E(& \bx, \bbv) = - \log \pi_{\bX}(\bx) - \tfrac{1}{2} \log |G(\bx)|  + \tfrac{1}{2} \bbv^\top G(\bx) \bbv, \notag \\[0.2cm]
%
\dot{\bx} &= \bbv, \label{eq:dynamics} \\
\dot{\bbv} &= 
- \dfrac{1}{2}G(\bx)^{-1}\big[ 2\partial_{\bx} G(\bx) - (\partial_{\bx}\textsl{vec} \ G(\bx))^\top\big] (\dot{\bx} \otimes \dot{\bx}) \notag \\
& - G(\bx)^{-1} \nabla \phi(\bx), \notag
\end{align}
where the notation $\partial_{\bx} A = [\partial_{x_1} A \cdots \partial_{x_D} A]$. Observe that the $k^{th}$ row of the matrix $\tfrac{1}{2}G(\bx)^{-1}\big[ 2\partial_{\bx} G(\bx) - (\partial_{\bx}\textsl{vec} \ G(\bx))^\top\big]$ is $(\textsl{vec} \ \Gamma^k)^\top$ where $\Gamma^{k}_{i, j}(\bx) =\tfrac{1}{2} \sum_l G^{k, l}(\frac{\partial}{\partial x_i} G_{i,j} + \frac{\partial}{\partial x_j} G_{i, l} - \frac{\partial}{\partial x_l} G_{i, j})$ are the Christoffel symbols \citep[See][for similar formulation]{arvanitidis2018latent}. The matrix elements $G_{i, j}$ and $G^{i, j}$ are the elements of the matrix $G(\bx)$ and its inverse respectively, and $\nabla \phi(\bx) = - \nabla \log \pi_{\bX}(\bx) + \frac{1}{2} \nabla \log \det G(\bx)$. 

The explicit integrator 
repeats $L_F$ times the updates
\begin{align}
\bbv^{(n + 1/2)} &= 
A_{n,n}^{-1}
\left[\bbv^{(n)} - \frac{\varepsilon}{2} G(\bx^{(n)})^{-1} \nabla \phi (\bx^{(n)})\right] \notag \\
\bx^{(n + 1)} &= \bx^{(n)} + \varepsilon \bbv^{(n + 1/2)} \label{eq:LMCupdates} \\
\bbv^{(n + 1)} &= 
A_{n+1, n+1/2}^{-1}
\Big[\bbv^{(n + 1/2)} \notag \\
 &\quad\quad\quad -\frac{\varepsilon}{2} G(\bx^{(n + 1)})^{-1} \nabla \phi (\bx^{(n + 1)})\Big] \notag
\end{align}
where $\Omega(\bx, \bbv)$ is a matrix whose $(i, j)$ element is given by  $\sum_k v_k \Gamma_{k, j}^i(\bx)$ and $A_{n_1,n_2} = I_D + \frac{\varepsilon}{2} \Omega(\bx^{(n_1)}, \bbv^{(n_2)})$.

The integrator is not volume-preserving and we need determinant adjustment for 
the acceptance probability
%
$
\alpha_{LMC} = \text{min} \left\lbrace 1, \exp \big( - E_{\text{diff}}
\big) |\det J| \right\rbrace  
$
where $E_{\text{diff}} = E(\bx^{(L_F + 1)}, \bbv^{(L_F + 1)})  - E(\bx^{(1)}, \bbv^{(1)})$ 
and
\begin{align*}
\det J &= \prod_{n = 1}^{L_F} 
\left ( 
\frac{
\det\big(G(\bx^{(n+1)}) - \frac{\varepsilon}{2} \tilde{\Omega}(\bx^{(n+1)}, \bbv^{(n+1)}) \big)
}{
\det\big( G(\bx^{(n+1)}) + \frac{\varepsilon}{2} \tilde{\Omega}(\bx^{(n+1)}, \bbv^{(n+1/2)})  \big)
}
\right .\\
&
\times
\left .
\frac{
\det\big( G(\bx^{(n)}) - \frac{\varepsilon}{2} \tilde{\Omega}(\bx^{(n)}, \bbv^{(n+1/2)}) \big)
}{
\det\big(G(\bx^{(n)}) + \frac{\varepsilon}{2} \tilde{\Omega}(\bx^{(n)}, \bbv^{(n)}) \big)
}
\right ),
\end{align*}
where the matrix $\tilde{\Omega}(\bx, \bbv) = G(\bx) \Omega(\bx, \bbv)$. 

LMC does not require fixed-point iterations and hence only needs two inversions per step,
but the computational advantage is lost due to computation of the determinants and the $\mathcal{O}(D^3)$ Christoffel symbols. The complexity of both RMHMC and LMC in a general metric is $\mathcal{O}(D^3)$, and their relative speed depends on the problem.


\paragraph{LMC in Monge Metric}

\begin{figure}
    \centering
    \includegraphics[width=0.9\columnwidth]{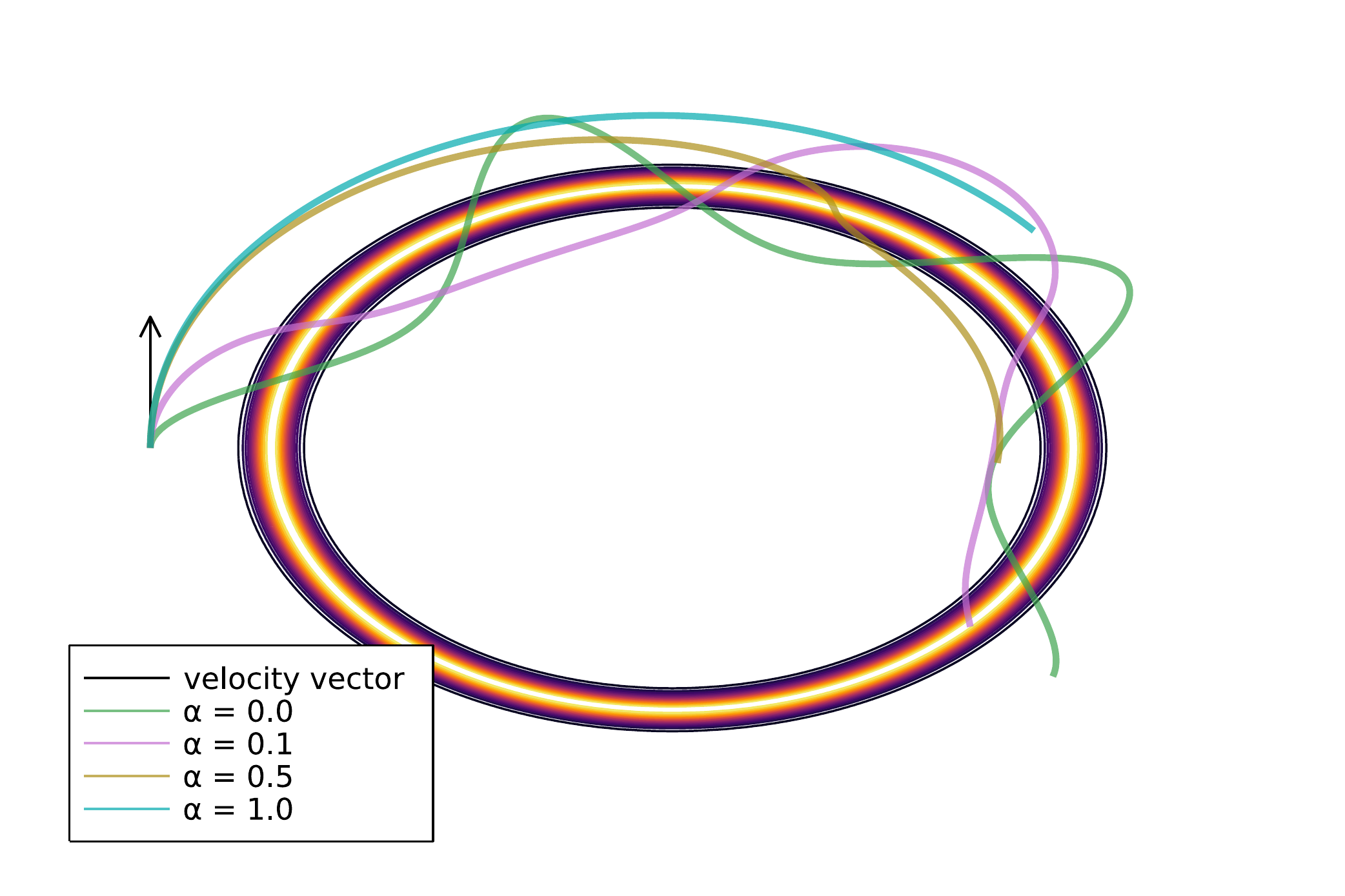}
    \caption{Geodesic paths of Lagrangian dynamics in Monge metrics of different $\alpha$ for fixed initial velocity. Note that $\alpha=0$ means Euclidean metric.}
    \label{fig:geodesics}
\end{figure}

In the Monge metric $G_M(\bx)$ the energy becomes
\begin{align*}
E(\bx, \bbv) &= - \ell(\bx) - \tfrac{1}{2} \log (1 + \alpha^2 \norm{\nabla \ell(\bx)}^2)  \nonumber \\
&\quad\quad + \tfrac{1}{2} \norm{\bbv}^2 + \tfrac{\alpha^2}{2} \left\langle  \nabla \ell(\bx),  \bbv \right\rangle^2.
\end{align*}
In the dynamical system \eqref{eq:dynamics}
we retain $\dot{\bx} = \bbv$ and for the velocity we have
%
\begin{align*}
\dot{\bbv} &= -\frac{\alpha^2}{1 + \alpha^2 \norm{\nabla \ell (\bx)}^2} \big( \nabla \ell (\bx) \ (\textsl{vec} \  H(\bx))^\top \big) (\bbv \otimes \bbv)  \\
& - G_M(\bx)^{-1} \left( \frac{\alpha^2 H(\bx)}{1 + \alpha^2 \norm{\nabla \ell(\bx)}^2} - I_D\right) \nabla \ell (\bx) .
 \end{align*}
%
%
%
For an initial velocity $\bbv$ and initial position $\bx$ these
keep the energy constant. The Hessian $H(\bx)$ appears here due to the Christoffel symbols, even though the metric only involves gradients.
We illustrate the geodesics for various $\alpha$ in Figure~\ref{fig:geodesics}, computed for a ring distribution. For $\alpha=0$ the metric reduces to Euclidean and the geodesic paths fluctuate around the typical set \citep[see][for detailed discussion]{betan:2017}, whereas for large $\alpha$ they resemble clear orbits.

After fairly extensive simplification, the update equations \eqref{eq:LMCupdates} in Monge metric can be written as in Table~\ref{tab:LMCe_updates}. The full derivation and a pseudo-code for the algorithm is provided in the Supplement. These updates are somewhat complicated, but free of matrix-matrix products and free of explicit matrix inversions. The proposal acceptance probability simplifies in a similar manner using
\begin{align*}
\det\Big( G(\bx) & \pm \frac{\varepsilon}{2} \tilde{\Omega}(\bx, \bbv) \Big) = \\
& 1 + \alpha^2 \norm{\nabla \ell(\bx)}^2 \pm \frac{\alpha^2 \varepsilon}{2} \langle \nabla \ell(\bx), H(\bx) \bbv \rangle .
\end{align*}
\begin{table*}
\caption{Numerical integration updates for LMC in the Monge metric. See Supplement for derivations.} \label{tab:LMCe_updates}
\begin{center}
\begin{minipage}{0.75\textwidth}
\begin{align*}
\bbv^{(n + 1/2)} &= \left[I_D - \frac{\nabla \ell(\bx^{(n)}) \big(\nabla \ell(\bx^{(n)})^\top + \frac{\varepsilon}{2}(\bbv^{(n)})^\top H(\bx^{(n)}) \big) }{\big[\nabla \ell(\bx^{(n)})^\top + \frac{\varepsilon}{2}(\bbv^{(n)})^\top H(\bx^{(n)}) \big]^\top \nabla \ell(\bx^{(n)}) + \tfrac{1}{\alpha^2}} \right] \\ & \times \left\lbrace \left[\Big( \alpha^2\nabla \ell(\bx^{(n)})^\top \bbv^{(n)} + \frac{\varepsilon}{2} \Big) I_D - \frac{\varepsilon \alpha^2}{2 + 2\alpha^2 \norm{\nabla \ell(\bx^{(n)})}^2} H(\bx^{(n)}) \right] \nabla \ell(\bx^{(n)})  + \bbv^{(n)} \right\rbrace \\
\bx^{(n + 1)} &= \bx^{(n)} + \varepsilon \bbv^{(n + 1/2)} \\
\bbv^{(n + 1)} &= \left[I_D - \frac{\nabla \ell(\bx^{(n + 1)}) \big(\nabla \ell(\bx^{(n + 1)})^\top + \frac{\varepsilon}{2}(\bbv^{(n + 1/2)})^\top H(\bx^{(n + 1)}) \big) }{\big[\nabla \ell(\bx^{(n + 1)})^\top + \frac{\varepsilon}{2}(\bbv^{(n + 1/2)})^\top H(\bx^{(n + 1)}) \big]^\top \nabla \ell(\bx^{(n + 1)}) + \tfrac{1}{\alpha^2}} \right] \\ & \times \left\lbrace \left[ \Big( \alpha^2 \nabla \ell(\bx^{(n + 1)})^\top \bbv^{(n + 1/2)} + \frac{\varepsilon}{2} \Big) I_D - \frac{\varepsilon \alpha^2}{2 + 2 \alpha^2 \norm{\nabla \ell(\bx^{(n+1)})}^2} H(\bx^{(n + 1)}) \right] \nabla \ell(\bx^{(n  + 1)})  + \bbv^{(n + 1/2)} \right\rbrace
\end{align*}
\end{minipage}
\end{center}
\end{table*}

The computation for one pass of the numerical integrator is dominated by the formation of the gradient vector ($\mathcal{O}(D)$) and the Hessian matrix ($\mathcal{O}(D^2)$). Since they are called twice in each loop of the numerical integrator, the cost is dominated by $2 L_F (\mathcal{O}(D) + \mathcal{O}(D^2))$ operations. The overall complexity is hence quadratic in $D$, not cubic as with the Fisher metric.

\section{EXPERIMENTS}

We evaluate LMC in Monge metric (LMC-Monge) in two example problems, a funnel distribution and posterior inference for logistic regression, but note that Figures~\ref{fig:illustration} and \ref{fig:geodesics} already demonstrated the metric in two other contexts.
We compare against competing methods in Euclidean and Fisher metrics (when applicable).
The experiments were ran on Intel i5-8250@1.6GHz laptop CPU. All experimental details and some additional illustrations are provided in the Supplement. The methods were implemented in {\tt Julia} \citep{julia} and the implementation is available at {\tt https://github.com/mahaa2/EmbeddedLMC}, providing both the inference algorithm itself as well as scripts for re-creating some of the experiments.

\subsection{Funnel Distribution}

We first show the metric helps in exploring areas of strong curvature, using the funnel distrubution by \citet{neal:2003}.
The $D$-dimensional funnel is given by
\begin{equation}
 \pi_{\bX}(\bx, a) = \prod_{i = 1}^{D} \mathcal{N}(x_i|0, \text{softplus}(a))\mathcal{N}(a|\mu, \sigma^2_a),
\end{equation}
where the marginal distribution of $a$ is  $\mathcal{N}(a|\mu, \sigma^2_{a})$ and hence we can easily evaluate the quality of the marginal. We set $\mu = 0.0$ and $\sigma^2_{a} = 15.0$, and use $60.000$ samples. To illustrate the metric we use $\alpha=1$ with accurate numeric integration with small step-length $\epsilon$ and $L_F$ growing from $8$ to $130$ when increasing the dimensionality, adjusted by visual inspection.

\begin{figure}[t]
\begin{center}
\includegraphics[width=\columnwidth]{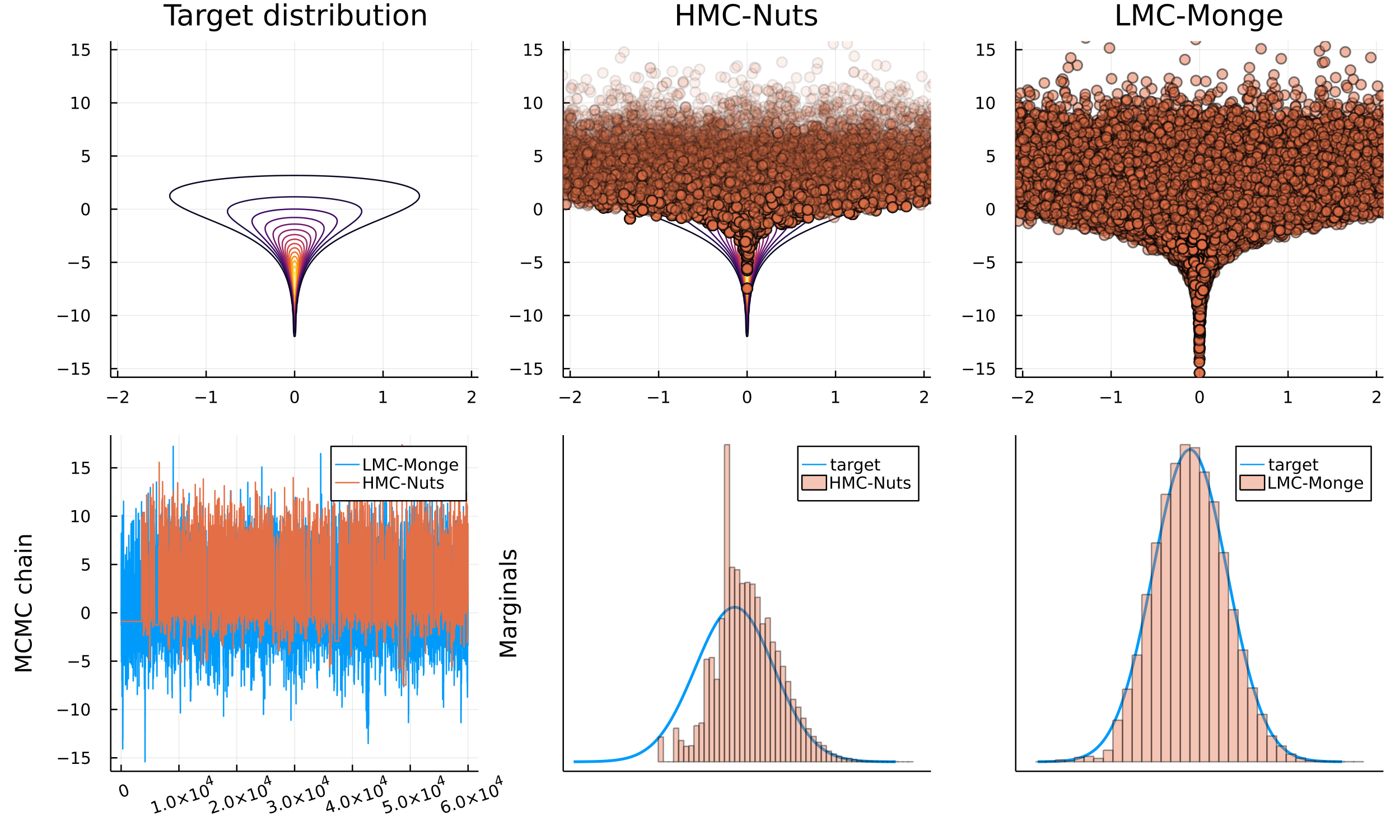} 
\end{center}
\caption{1D funnel. LMC in Monge metric (right) explores the target distribution well, whereas HMC in Euclidean metric (middle) does not, even though both chains mix well (bottom left).}
\label{fig:funnel1}
\end{figure}

Figure~\ref{fig:funnel1} demonstrates how LMC-Monge provides samples from the correct distribution but HMC in Euclidean metric does not, even when using the more advanced NUTS algorithm \citep{hoffman:14} as implemented in \texttt{Turing.jl} \citep{turing}. Both samplers have low autocorrelation, seen by observing the sampling chains, and hence the problems of the Euclidean sampler could easily go unnoticed in practice. Fisher metric is here not applicable since we are not conducting posterior inference, but rather sampling from the distribution itself for given parameters.

\begin{figure}[t]
\begin{center}
\includegraphics[width = 0.9\columnwidth]{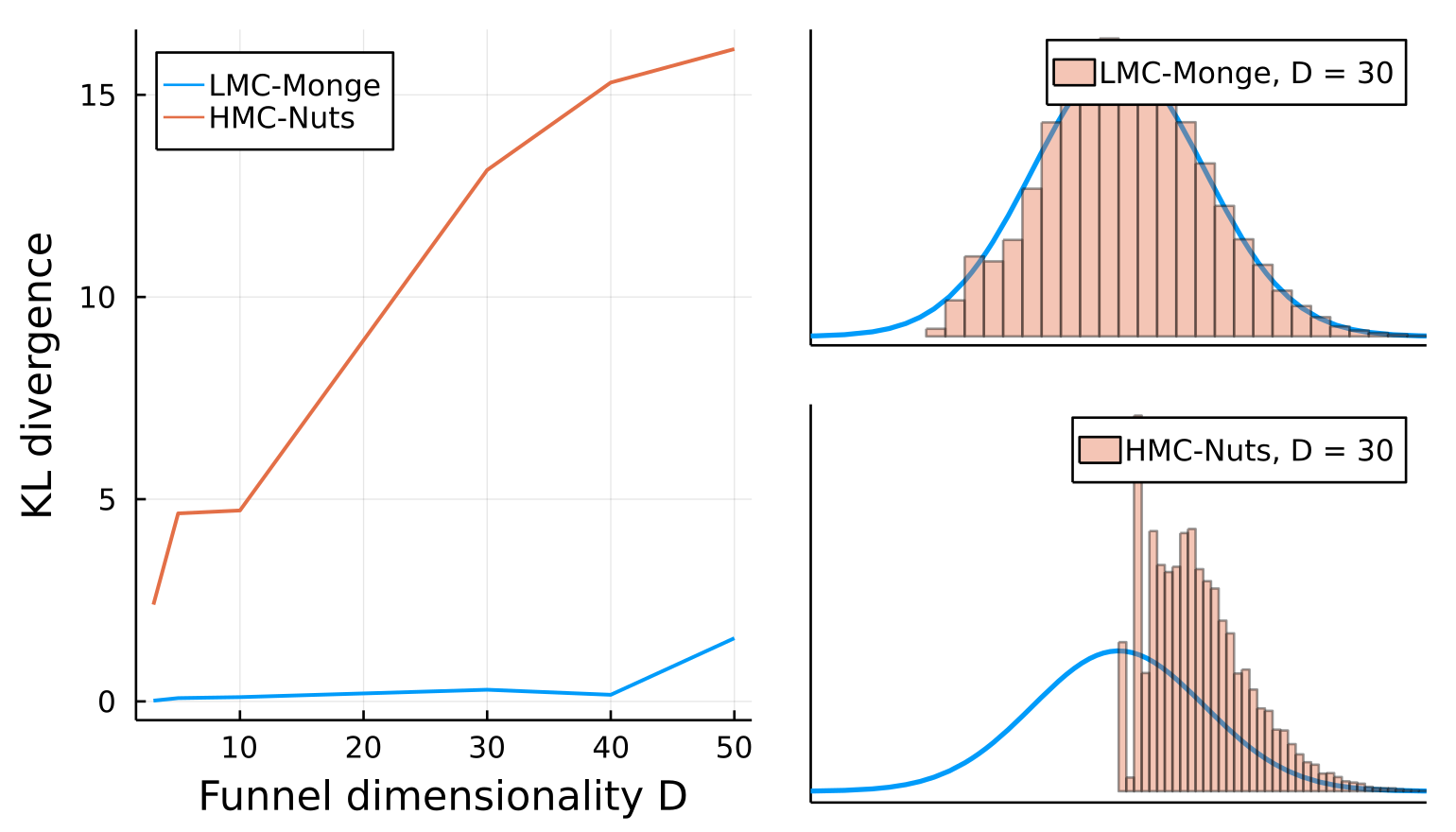} 
\end{center}
\caption{(Left) KL divergence between the true marginal $\mathcal{N}(a)$ and its estimate as a function of problem dimensionality. (Right) Marginals for $D=30$.}
\label{fig:funnel2}
\end{figure}

Figure~\ref{fig:funnel2} investigates the quality as a function of $D$, measured by approximating KL divergence between the true marginal and the MCMC approximation with $\sum_k [\log P(A_k) / \tilde{Q}(A_k)] P(A_k)$, where $A_k$ is a histogram bin. LMC-Monge retains good accuracy for all $D$ whereas NUTS gets progressive worse. 
%
\subsection{Logistic Regression}

Having established the metric can explore well, we turn the attention to performance. We replicate the logistic regression experiment of \citet{lan:2015} on their largest data sets, using 20,000 samples (warm-up of 5,000). We also otherwise match their empirical setup, and in particular select $\epsilon$ and $L_F$ to obtain acceptance probability in the range of 0.6-0.9 for each method. We evaluate the efficiency using the standard effective sample size (ESS) measure, but note that high ESS does not guarantee correct sampling. 
%

Table~\ref{tab:lmcexp} compares LMC-Monge against three baselines (using implementation and parameter settings of \citet{lan:2015}) that differ in terms of the metric: LMC and RHMC in Fisher metrics, and standard HMC in spherical Euclidean metric. The Riemannian methods have clearly higher ESS compared to the Euclidean HMC, and Monge metric behaves similarly to the Fisher metric but is faster. This validates our main claim. For completeness, we also show the results for NUTS in Euclidean metric even though direct comparison is not fair due to adaptive choice of $L_F$ and $\epsilon$ that also helps in achieving high ESS. 

\begin{figure}[t]
    \centering
    \includegraphics[width = \columnwidth]{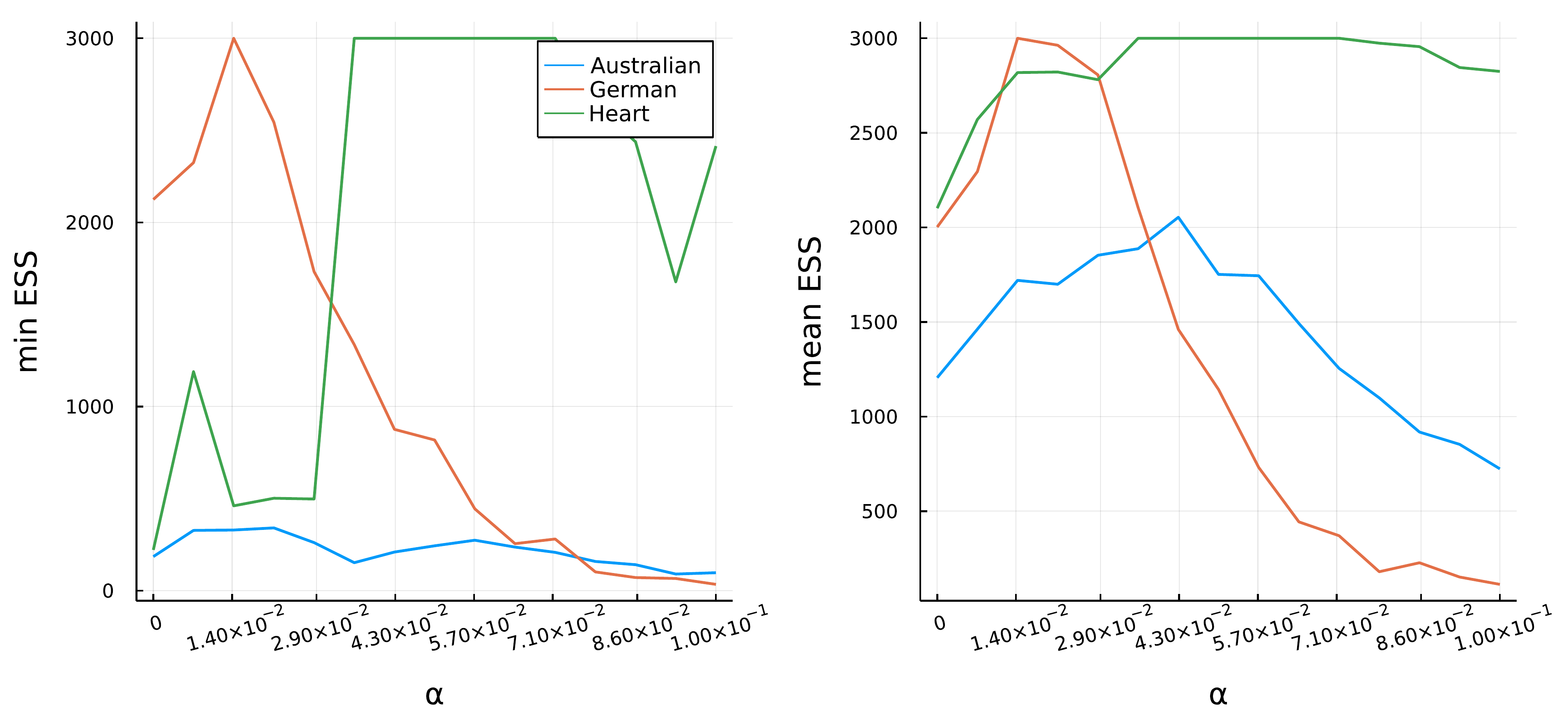} 
    \caption{Relationship between $\alpha$ and sample efficiency. The optimal $\alpha$ depends on the problem.}
    \label{fig:alphatest}
\end{figure}

Figure~\ref{fig:alphatest} shows the effect of the control parameter, using 3000 samples after warm-up of 500. The optimal choice depends on the data and often very small $\alpha$ are best, but we note that this does not necessarily mean the metric would be particularly close to Euclidean as the magnitude of $\nabla \ell(\bx)$ and $H(\bx)$ can also be large.

\begin{table*}[t]
\caption{Logistic regression. The best method (ESS/sec) with constant $L_F$ is indicated by boldface, and boldface italics marks cases where NUTS with adaptive $L_F$ is overall the best.
AP is average acceptance probability.}
\label{tab:lmcexp}
{\small
\begin{tabular}{llrrrrr}
Data                 & Method       & AP & ESS (min, mean, median)       & time(s) & min(ESS)/s     & mean(ESS)/s  \\ \hline\hline
Heart                & LMC-Monge ($\alpha=0.01$) & $0.79$ & $(15000, 15000, 15000)$ & $34$ & ${\bf 441}$ & ${\bf 441}$ \\
$N = 270$            & LMC-Fisher   & $0.76$ & $(10347, 10848, 10724)$ & $63$  & $164$        & $172$        \\
$D = 14$             & RMHMC-Fisher & $0.72$ & $(6263,  7391,   7430)$ & $90$  & $69$         & $82$     \\
                     & HMC-Euclidean          & $0.71$ & $(378, 1164, 2624)$     & $7$   & $55$   & $170$ \\
                     & {\it HMC-Nuts}     & ${\it 0.94}$ & ${\it (13804, 14777, 15000)}$ & ${\it 15}$  & \textbf{\textit{927}}  & \textbf{\textit{1055}} \\
\hline
German               & LMC-Monge ($\alpha=0.01$)   & $0.81$ & $(13390, 14949, 15000)$ & $71$ & ${\bf 194}$ & ${\bf 210}$\\ 
$N = 1000$           & LMC-Fisher   & $0.70$ & $(13762, 14932, 15000)$ & $202$ & $68$        & $74$   \\
$D = 22$             & RMHMC-Fisher & $0.75$ & $(14885, 14995, 15000)$ & $252$ & $49$        & $59$ \\
                     & HMC-Euclidean        & $0.73$ & $(766, 4803, 15000)$    & $69$  & $11$        & $69$ \\
                     & {\it HMC-Nuts}     & ${\it 0.70}$ & ${\it (14168, 14960, 15000)}$ & ${\it 40}$  &  \textbf{\textit{350}} & \textbf{\textit{374}}  \\
\hline                     
Australian           & LMC-Monge ($\alpha=0.01$)   & $0.82$ & $(1259, 12932, 15000)$  & $52$ & $24$        & ${\bf 249}$ \\ 
$N = 690$            & LMC-Fisher   & $0.75$ & $(9636, 10464, 10443)$   & $100$ & ${\bf 96}$  & $104$ \\
$D = 15$             & RMHMC-Fisher & $0.72$ & $(7824, 9237,   9055) $  & $134$ & $58$        & $69$  \\
                     & HMC-Euclidean          & $0.74$ & $(1225, 4440, 10691)$    & $18$  & $65$        & $246$ \\
                     & {\it HMC-Nuts}     & ${\it 0.99}$ & ${\it (1227, 11715, 15000)}$   & ${\it 54}$  & ${\it 22}$        & ${\it 216}$  \\
\hline
\end{tabular}%
}
\end{table*}

\section{DISCUSSION}

Augmented MCMC is the workhorse of probabilistic programming. Geometric MCMC algorithms offer theoretical advantages for complex distribution,
but have slow updates and are rarely used in practice. We set out to resolve this problem, by providing a new Riemannian metric that still accounts for local curvature but is faster.
The Monge metric, a natural metric for a Monge patch embedding, can be easily computed for every density based on gradients alone and has efficient inverse and determinant. Besides LMC, it could be used e.g. with the explicitly symplectic integrator for RMHMC \citep{cobb:2019} or with manifold-adjusted Langevin Monte Carlo \citep{girolami:2011}.

We demonstrated the basic properties of the metric,
but significant practical steps remain on the path to validating it in routine use. A practical tool for probabilistic programming or arbitrary sampling tasks would require a high-quality implementation and automatic means for adapting the controls parameter $\epsilon$, $L_F$ and $\alpha$. We expect extension of the NUTS to Riemannian metrics \citep{betan:2013} to help by offering automatic choice of the integration length, and $\alpha$ could be adapted during warm-up similar to how the Euclidean metric tensor $M$ is often adapted, for instance based on gradient magnitudes. Finally, we see a need for more detailed theoretical analysis of the metric, e.g. along the lines of \citet{brosse:2018}.




\section*{Acknowledgments}

Hartmann and Klami were supported by the Academy of Finland (grant 345811 and Flagship programme: Finnish Center for Artificial Intelligence, FCAI), and Business Finland (MINERAL project). 

Mark Girolami was supported by Engineering and Physical Sciences Research Council Grants [EP/R034710/1, EP/R018413/1, EP/R004889/1, EP/P020720/1] and a Royal Academy of Engineering Research Chair.

We thank ST John for software tips and Luiz Hartmann for discussions on topics in differential geometry.

\bibliographystyle{plainnat}
\bibliography{refs}


\clearpage
\appendix

\thispagestyle{empty}

\onecolumn \makesupplementtitle

\section{OVERVIEW}

This Supplementary material provides additional derivations and details for the article \emph{Lagrangian Manifold Monte Carlo on Monge Patches}. Sections~\ref{sec:derivations}, \ref{sec:integrator} and \ref{sec:squareroot} provide the derivations to complement Sections 3 and 4 of the main paper, whereas Section~\ref{sec:experiments} provides the full experimental details, additional result plots and experiments. 

\section{DERIVATIONS AND ADDITIONAL FORMULATIONS}
\label{sec:derivations}

In this section, we present derivations and mathematical simplifications that verify statements provided in the main paper and that are required for derivation of the LMC Monge update rules provided in Section~\ref{sec:integrator}. Throughout this section, we consider shortened notation $L_\alpha(\bx) = 1 + \alpha^2 \norm{\nabla \ell(\bx)}^2$ whenever convenient. 


\paragraph{Christoffel symbols}
Section~3.2 provided a compact closed-form expression for the Christoffel symbols in the Monge metric. Starting with the formal definition of a $D$-dimensional manifold and particularizing for the case of our proposed embedding, we thus have
\begin{align}
\Gamma^k_{i, j}(\bx) &= \sum_{l = 1}^D G^{-1}_{k, l} (\bx) \left\langle \frac{\partial^2}{\partial x_i \partial x_j}  \hspace{0.03cm} \Xi, \frac{\partial}{\partial x_l} \hspace{0.03cm} \Xi \right\rangle 
= \alpha^2 \sum_{l = 1}^D G^{-1}_{k, l}(\bx) \ \frac{\partial}{\partial x_l} \ell(\bx) \frac{\partial^2}{\partial x_i \partial x_j}  \ell (\bx) \nonumber \\
&= \alpha^2 \sum_{l = 1}^D \left( \delta_{k, l} - \alpha^2 \frac{ \ \frac{\partial}{\partial x_k} \ell(\bx)  \frac{\partial}{\partial x_l} \ell(\bx)}{1 + \alpha^2 \norm{\nabla \ell(\bx)}^2}\right) \frac{\partial}{\partial x_l} \ell(\bx) \frac{\partial^2}{\partial x_i \partial x_j}  \ell (\bx) \nonumber \\
&= \alpha^2 \left(1 - \alpha^2 \frac{\norm{\nabla \ell(\bx)}^2 }{1 + \alpha^2 \norm{\nabla \ell(\bx)}^2} \right) \frac{\partial}{\partial x_k} \ell(\bx) \frac{\partial^2}{\partial x_i \partial x_j}  \ell (\bx) \nonumber \\
& = \frac{\alpha^2}{1 + \alpha^2\norm{\nabla \ell(\bx)}^2} \frac{\partial}{\partial x_k} \ell (\bx) \frac{\partial^2}{\partial x_i \partial x_j}  \ell (\bx), \nonumber
\end{align}
which corresponds to the expression provided in the main paper. Since the Christoffel symbols are symmetric over the indices $i, j$, we can further express them in full matrices as
$$\Gamma^{k} = \frac{\alpha^2}{1 + \alpha^2\norm{\nabla \ell(\bx)}^2} \frac{\partial}{\partial x_k} \ell (\bx) \ H(\bx)$$ for $k = 1, \ldots, D$, where $H(\bx) = \nabla \nabla \ell(\bx)$ is the Hessian matrix of the log target distribution.

\paragraph{Matrix $\Omega(\bx, \bbv)$}
The LMC updates (Eq. (3) in main paper) depend on the matrix $\Omega(\bx, \bbv)$, which is a matrix whose $(i, j)$ element is given by $\sum_k v_k \Gamma_{k, j}^i(\bx)$. In full matrix form this simplifies to
\begin{align*} 
\Omega &= \begin{bmatrix}
\bbv^\top \Gamma(\bx)_{\bullet, 1}^{1} & \cdots & \bbv^\top \Gamma(\bx)_{\bullet, D}^{1} \\ 
\vdots & \ddots & \vdots \\ 
\bbv^\top \Gamma(\bx)_{\bullet, 1}^{D} & \cdots & \bbv^\top \Gamma(\bx)_{\bullet, D}^{D}
\end{bmatrix}
= (I_D \otimes \bbv^\top) \begin{bmatrix}
\Gamma^1(\bx) \\ 
\vdots \\ 
\Gamma^D(\bx)
\end{bmatrix} 
= (I_D \otimes \bbv^\top) \frac{\alpha^2}{L_\alpha(\bx)} \begin{bmatrix}
\frac{\partial}{\partial x_1} \ell (\bx) H(\bx) \\ 
\vdots \\ 
\frac{\partial}{\partial x_D} \ell (\bx) H(\bx) 
\end{bmatrix} \\
& =\frac{\alpha^2}{L_\alpha(\bx)}(I_n \otimes \bbv^\top) (\nabla \ell \otimes H(\bx))
=   \frac{\alpha^2}{L_\alpha(\bx)} \nabla \ell(\bx) \otimes (\bbv^\top H(\bx)) = \frac{\alpha^2}{L_\alpha(\bx)} \nabla \ell(\bx) \ (\bbv^\top H(\bx)).
\end{align*}
Building on this, the matrix $\tilde{\Omega}(\bx, \bbv) = G(\bx) \ \Omega(\bx, \bbv)$ required for the determinant adjustment and the inverses in the numerical integrator updates reduces to
\begin{align*}
\tilde{\Omega}(\bx, \bbv) &= \big(I_D + \alpha^2 \nabla \ell(\bx) \nabla \ell(\bx)^\top \big) \tfrac{\alpha^2}{L_\alpha(\bx)} \nabla \ell(\bx) \  (\bbv^\top H(\bx)) \\ 
&= \alpha^2 \left( \frac{L_\alpha(\bx) - 1}{L_\alpha(\bx)} \nabla \ell(\bx) \ (\bbv^\top H(\bx))  + \frac{1}{L_\alpha(\bx)} \nabla \ell(\bx) \ (\bbv^\top H(\bx)) \right) \\
&= \alpha^2 \ \nabla \ell(\bx) \ (\bbv^\top H(\bx)).
\end{align*}
Both of these will be required for simplifying the update rules in the Monge metric.

\paragraph{Determinant}
For the determinant $\det \big(G((\bx) + \tfrac{\varepsilon}{2} \tilde{\Omega}(\bx, \bbv) \big)$, that is necessary in the Metropolis-Hasting acceptance probability rule, we use the Sherman-Morrison matrix lemma to get 
\begin{align}
\det G((\bx) \pm \tfrac{\varepsilon}{2} \tilde{\Omega}(\bx, \bbv) &= \det G(\bx) \det I_D \pm \tfrac{\varepsilon}{2} \Omega(\bx, \bbv) \notag \\ 
&= L_\alpha(\bx) \det I_D \pm \tfrac{\varepsilon}{2} \tfrac{\alpha^2}{L_\alpha(\bx)} \nabla \ell(\bx) \bbv^\top H \notag \\
&= L_\alpha(\bx)  \big( 1 \pm \tfrac{\varepsilon}{2} \tfrac{\alpha^2}{L_\alpha(\bx)} \langle \nabla \ell, H(\bx) \bbv \rangle \big) \notag \\
& = 1 + \alpha^2 \norm{\nabla \ell(\bx)}^2 \pm \tfrac{\varepsilon}{2} \alpha^2 \langle \nabla \ell(\bx), H(\bx) \bbv \rangle. \label{eq:det}
\end{align}

\paragraph{Inverses}
The inverse matrices required in the first and third updating equation of the velocity vector of the explicitly numerical integrator are also simplified using the same matrix lemma. We have
\begin{align} \label{eq:inv}
\left(G(\bx) +\tfrac{\varepsilon}{2} \tilde{\Omega}(\bx, \bbv)\right)^{-1} &= \left(I_D + \alpha^2 \ \nabla \ell (\bx) \nabla \ell (\bx)^\top + \frac{\varepsilon \alpha^2}{2} \nabla \ell(\bx) \  (\bbv^\top H(\bx)) \right)^{-1} \notag \\
&= \Big(I_D + \alpha^2 \nabla \ell(\bx) \big(\nabla \ell(\bx)^\top + \tfrac{\varepsilon}{2}\bbv^\top H(\bx) \big) \Big)^{-1} \notag \\
& = I_D -  \frac{\nabla \ell(\bx) \ \big(\nabla \ell(\bx)^\top + \tfrac{\varepsilon}{2} \bbv^\top H(\bx) \big)}{\big(\nabla \ell(\bx)^\top + \tfrac{\varepsilon}{2} \bbv^\top H(\bx) \big)\nabla \ell(\bx) + \frac{1 }{\alpha^2}}. 
\end{align}

\paragraph{Gradient of potential energy}
The potential energy for LMC is $\phi(\bx) = - \log \pi_{\bX}(\bx) + \frac{1}{2} \log \det G(\bx)$ and we need the gradient of that. The first term is obvious and the latter can be computed using
%
\begin{align*}
 \dfrac{\partial}{\partial x_i} \log & \det G(\bx) = tr \left( G^{-1}(\bx) \dfrac{\partial}{\partial x_i} G(\bx) \right) \notag \\
  &= \alpha^2 \ tr \left[ \left( I_D -  \alpha^2 \dfrac{\nabla \ell(\bx) \ \nabla \ell(\bx)^\top}{L_\alpha(\bx)} \right) 2 \left(\dfrac{\partial}{\partial x_i} \nabla \ell(\bx) \right) \nabla \ell(\bx)^\top \right] \notag \\
 &= \frac{2\alpha^2}{L_\alpha(\bx)} \ \big[ tr (H_i(\bx) \nabla \ell(\bx)^\top L_\alpha(\bx))  -  \alpha^2 tr (\nabla \ell(\bx) \ \nabla \ell(\bx)^\top H_i(\bx) \nabla \ell(\bx)^\top )  \big] \notag \\
&= 2 \alpha^2 \nabla \ell(\bx)^\top H_i(\bx) \Big( 1 - \alpha^2 \frac{\nabla \ell(\bx)^\top \nabla \ell(\bx)}{L_\alpha(\bx)} \Big) 
= 2 \alpha^2 \Big( 1 - \alpha^2 \frac{\nabla \ell(\bx)^\top \nabla \ell(\bx)}{L_\alpha(\bx)} \Big) \nabla \ell(\bx)^\top H_i(\bx) \notag \\
&= 2 \alpha^2 \frac{\nabla \ell(\bx)^\top}{L_\alpha(\bx)} H_i(\bx),
\end{align*}
where $H_i(\bx)$ is the $i^{th}$ row (or column) of the Hessian matrix. Hence the gradient of the second term in the potential energy is given by
 \begin{align}\label{eq:gradp}
\nabla \log \det G =  \frac{2\alpha^2}{L_\alpha(\bx)} H(\bx) \ \nabla \ell(\bx).
\end{align}

\paragraph{Relationship to Fisher metric}

Section~3.1 established a relationship between Fisher and Monge metrics. For the specific case where $\ell(\bx) = \log \pi_Y (Y|\bx)$ corresponds to a data generating distribution over some random variable $Y$, we can compute the expectation of the Monge metric over $Y$. We can then write
\[
\Ex_Y\left(G_M(\bx)\right) = I_D + \alpha^2 \Ex_Y\left (\nabla \ell(\bx) \nabla \ell(\bx)^\top \right ),
\]
where the latter term corresponds to the definition of Fisher metric $G_F(\bx)$ as expressed in Section~2.
Consequently, we can interpret the expected Monge metric as a biased (and scaled) estimator for the Fisher metric, for instance writing
\[
\Ex_Y\left(G_M(\bx)\right) = I_D + \alpha^2 G_F(\bx)
\]
or equivalently
\[
G_F(\bx) = \alpha^{-2} \left [ \Ex_Y\left(G_M(\bx)\right) - I_D \right ].
\]

\section{CLOSED-FORM EXPLICIT NUMERICAL INTEGRATOR}
\label{sec:integrator}

Having established the required computational elements in the previous section, we proceed to derivation of the update rules for the explicit numerical integrator for LMC-Monge.
Full pseudo-code for the resulting algorithm is given in Algorithm~\ref{alg:pseudocode} and reference implementation in {\tt Julia} is provided at
${\tt https://github.com/mahaa2/EmbeddedLMC}$. The code also includes scripts for re-creating some of the experiments.


The LMC algorithm assumes that step-size $\varepsilon$, number of steps $L_F$ and the metric control parameter $\alpha$ are provided as inputs, together with some initial value for $\bx^{(1)}$ (the previous sample). The integrator proposed by \citet{lan:2015} then repeats the following steps $L_F$ times:
%
\begin{align}
\bbv^{(n + 1/2)} &= 
\Big[ G(\bx^{(n)}) + \frac{\varepsilon}{2} \tilde{\Omega}(\bx^{(n)}, \bbv^{(n)}) \Big]^{-1}
\left[G(\bx^{(n)}) \bbv^{(n)} - \frac{\varepsilon}{2} \nabla \phi (\bx^{(n)})\right] \\
\bx^{(n + 1)} &= \bx^{(n)} + \varepsilon \bbv^{(n + 1/2)} \notag \\
\bbv^{(n + 1)} &= 
\Big[ G(\bx^{(n+1)}) + \frac{\varepsilon}{2} \tilde{\Omega}(\bx^{(n+1)}, \bbv^{(n+1/2)}) \Big]^{-1}
\Big[ G(\bx^{(n + 1)})^{-1} \bbv^{(n + 1/2)}
  - \frac{\varepsilon}{2} \nabla \phi (\bx^{(n + 1)})\Big]. \notag
\end{align}

To express these updates in the Monge metric, we will use the simplifications described in the previous section. The update for the position $\bx$ does not depend on the metric, whereas the two updates for the velocity $\bbv$ are analogous, requiring the same algebraic changes. Consequently, we only write the first update explicitly, using \eqref{eq:gradp} to compute the gradient of the energy and \eqref{eq:inv} to compute the inverse. This results the expression provided also in Table~1 of the main paper:
%
\begin{align*}
\bbv^{(n + 1/2)} &= \Big[ G(\bx^{(n)}) + \frac{\varepsilon}{2} \tilde{\Omega}(\bx^{(n)}, \bbv^{(n)}) \Big]^{-1} \Big[G(\bx^{(n)}) \bbv^{(n)} - \frac{\varepsilon}{2} \nabla \phi (\bx^{(n)})\Big] \\
  &= \bigg[ I_D - \frac{\nabla \ell(\bx^{(n)}) \big(\nabla \ell(\bx^{(n)})^\top + \frac{\varepsilon}{2}(\bbv^{(n)})^\top H(\bx^{(n)}) \big) }{\big[\nabla \ell(\bx^{(n)})^\top + \frac{\varepsilon}{2}(\bbv^{(n)})^\top H(\bx^{(n)}) \big]^\top \nabla \ell(\bx^{(n)}) + \tfrac{1}{\alpha}} \bigg] \\ 
  & \hspace{4.0cm} \times 
  \bigg \lbrace \bigg[ \big( \alpha^2 \nabla \ell(\bx^{(n)})^\top \bbv^{(n)} + \frac{\varepsilon}{2} \big) I_D
   - \frac{\alpha^2 \varepsilon}{2L_\alpha} H(\bx^{(n)}) \bigg] \nabla \ell(\bx^{(n)}) + \bbv^{(n)} \bigg \rbrace.
\end{align*}

The integrator is not volume-preserving \citep[see][]{lan:2015}. Thus the proposal' acceptance probability needs the determinant adjustment and becomes
\begin{align*}
\alpha_{LMC} = & \text{min} \Big \lbrace 1, \exp \big( - E(\bx^{(L_F + 1)}, \bbv^{(L_F + 1)})
+ E(\bx^{(1)}, \bbv^{(1)})  \big) |\det J|  \Big \rbrace
\end{align*}
where the energy function $E$ is defined as,
\begin{align*}
E(\bx, \bbv) &= - \ell(\bx) - \tfrac{1}{2} \log \det G(\bx)  + \tfrac{1}{2} \bbv^\top G(\bx) \bbv \\
&= - \ell(\bx) - \tfrac{1}{2} \log (1 + \alpha^2 \norm{\nabla \ell(\bx)}^2) + \tfrac{1}{2} \norm{\bbv}^2
+ \frac{\alpha^2}{2} \left\langle  \nabla \ell(\bx),  \bbv \right\rangle^2
\end{align*}
and the determinant adjustment becomes
\begin{align*}
\det J &= \prod_{n = 1}^{L_F} 
\frac{
\det\big(G(\bx^{(n+1)}) - \frac{\varepsilon}{2} \tilde{\Omega}(\bx^{(n+1)}, \bbv^{(n+1)}) \big)
}{
\det\big( G(\bx^{(n+1)}) + \frac{\varepsilon}{2} \tilde{\Omega}(\bx^{(n+1)}, \bbv^{(n+1/2)})  \big)
}
\frac{
\det\big( G(\bx^{(n)}) - \frac{\varepsilon}{2} \tilde{\Omega}(\bx^{(n)}, \bbv^{(n+1/2)}) \big)
}{
\det\big(G(\bx^{(n)}) + \frac{\varepsilon}{2} \tilde{\Omega}(\bx^{(n)}, \bbv^{(n)}) \big)
}
\\
&= \prod_{n = 1}^{L_F} 
\frac{ L_\alpha(\bx^{(n+1)}) - \frac{\alpha^2 \varepsilon}{2} \langle \nabla \ell(\bx^{(n+1)}), H(\bx^{(n+1)}) \bbv^{(n+1)} \rangle
}{
L_\alpha(\bx^{(n+1)}) + \frac{\alpha^2 \varepsilon}{2} \langle \nabla \ell(\bx^{(n+1)}), H(\bx^{(n+1)}) \bbv^{(n+1/2)} \rangle
}
\frac{
L_\alpha(\bx^{(n)}) - \frac{\alpha^2 \varepsilon}{2} \langle \nabla \ell(\bx^{(n)}), H(\bx^{(n)}) \bbv^{(n+1/2)} \rangle
}{
L_\alpha(\bx^{(n)}) + \frac{\alpha^2 \varepsilon}{2} \langle \nabla \ell(\bx^{(n)}), H(\bx^{(n)}) \bbv^{(n)} \rangle,
}
\end{align*}
using the simplification provided in \eqref{eq:det}.


\begin{center}
\begin{algorithm*}
\SetAlgoLined
\KwResult{A sample from the distribution $\pi_{\bX}(\cdot)$}
 Inputs : $\nabla \ell$, $H$, $\varepsilon$, $L_F$, $\alpha$ and $\bx^{(1)}$\;
  Sample new velocity $\bbv^{(1)} = \sqrt{G^{-1}(\bx^{(1)})} \boldsymbol{z}$ where $z \sim \mathcal{N} \left(0, I_D \right)$\;
 Calculate current $E_1 = E(\bx^{(1)}, \bbv^{(1)})$\;
 $\Delta \log \det = 0$\;
 \For{$n = 1, \ldots, L_F$}{
 $\Delta \log \det =\Delta \log \det - \log\big|(1 + \frac{\varepsilon}{2L_\alpha(\bx^{(n)})} \nabla \ell(\bx^{(n)})^\top H(\bx^{(n)}) \bbv^{(n)})\big|$\;
 \# update velocity explicitly with a half-step\;
 \vspace{-0.6cm}
 \begin{align*}
 \bbv^{(n + 1/2)} &= \left[I_D - \frac{\nabla \ell(\bx^{(n)}) \big(\nabla \ell(\bx^{(n)})^\top + \frac{\varepsilon}{2}(\bbv^{(n)})^\top H(\bx^{(n)}) \big) }{\big[\nabla \ell(\bx^{(n)})^\top + \frac{\varepsilon}{2}(\bbv^{(n)})^\top H(\bx^{(n)}) \big]^\top \nabla \ell(\bx^{(n)}) + \tfrac{1}{\alpha^2}} \right] \\
  & \ \phantom{=} \times \left\lbrace \left[ \big( \alpha^2 \nabla \ell(\bx^{(n)})^\top \bbv^{(n)} + \frac{\varepsilon}{2} \big) I_D - \frac{\alpha^2 \varepsilon}{2L_{\alpha}(\bx^{(n)})} H(\bx^{(n)}) \right] \nabla \ell(\bx^{(n)})  + \bbv^{(n)} \right\rbrace
 \end{align*}
  $\Delta \log \det =\Delta \log \det + \log|(1 - \frac{\varepsilon}{2L_\alpha(\bx^{(n)})} \nabla \ell(\bx^{(n)})^\top H(\bx^{(n)}) \bbv^{(n+1/2)})|$\; 
  \# update position with a full-step\;
 \vspace{-0.6cm}
 \begin{align*}
  \bx^{(n + 1)} &= \bx^{(n)} + \varepsilon \bbv^{(n + 1/2)} 
 \end{align*}
   $\Delta \log \det =\Delta \log \det - \log\big|(1 + \frac{\alpha^2 \varepsilon}{2L_\alpha(\bx^{(n+1)})} \nabla \ell(\bx^{(n + 1)})^\top H(\bx^{(n + 1 )}) \bbv^{(n+1/2)}) \big|$\; 
    \# update velocity explicitly with a half-step\;
 \vspace{-0.6cm}
 \begin{align*}
 \bbv^{(n + 1)} &= \left[I_D - \frac{\nabla \ell(\bx^{(n + 1)}) \big(\nabla \ell(\bx^{(n + 1)})^\top + \frac{\varepsilon}{2}(\bbv^{(n + 1/2)})^\top H(\bx^{(n + 1)}) \big) }{\big[\nabla \ell(\bx^{(n + 1)})^\top + \frac{\varepsilon}{2}(\bbv^{(n + 1/2)})^\top H(\bx^{(n + 1)}) \big]^\top \nabla \ell(\bx^{(n + 1)}) + \tfrac{1}{\alpha^2}} \right] \\
  & \ \phantom{=} \times \left\lbrace \left[ \big( \alpha^2 \nabla \ell(\bx^{(n + 1)})^\top \bbv^{(n + 1/2)} + \frac{\varepsilon}{2} \big) I_D - \frac{\alpha^2 \varepsilon}{2L_{\alpha}(\bx^{(n)})} H(\bx^{(n + 1)}) \right] \nabla \ell(\bx^{(n + 1)})  + \bbv^{(n + 1/2)} \right\rbrace
 \end{align*} 
    $\Delta \log \det =\Delta \log \det + \log \big|(1 - \frac{\varepsilon}{2L_\alpha(\bx^{(n+1)})} \nabla \ell(\bx^{(n + 1)})^\top H(\bx^{(n + 1 )}) \bbv^{(n + 1)}) \big|$\; 
  }
   Calculate proposed $E_{L_F} = E(\bx^{(L_F + 1)}, \bbv^{(L_F + 1)})$\;
  Calculate $\textit{logRatio} = - E_1 + E_{L_F} + \Delta \log \det$\;
  Sample $u \sim U(0, 1)$\;
   \eIf{$\text{logRatio} > u$}{Accept $(\bx^{(L_F + 1)}, \bbv^{(L_F + 1)})$ as the current sample}{Reject $(\bx^{(L_F + 1)}, \bbv^{(L_F + 1)})$ and keep $(\bx^{(1)}, \bbv^{(1})$ as the current sample}
 \caption{Explicit Lagrangian Monte Carlo via embedding using the Monge patch}
 \label{alg:pseudocode}
\end{algorithm*}
\end{center}

\section{METRIC-TENSOR SQUARE ROOT AND VELOCITY SAMPLING}
\label{sec:squareroot}

To sample from the multivariate Gaussian $\bbv \sim \mathcal{N}(\0, G(\bx)^{-1})$, we need the square root matrix $\sqrt{G^{-1}(\bx)} = A$ such that $ G^{-1}(\bx) = AA^\top $. Since the inverse matrix of the metric is also formed by the outer-product of the gradients, it is possible to find the square root matrix with cost $\mathcal{O}(D^2)$, instead of the standard Cholesky decomposition with computational cost of $\mathcal{O}(D^3)$.

For inverse matrix of the metric-tensor $G_M(\bx)$, let the square root matrix be of the form $A = I_D + t  u u^\top$, where $t \in \mathbb{R}$. Then we have 
\begin{align*}
AA^\top = I_D + (2t + t^2 \norm{u}^2) u u^\top
\end{align*}
and we want that 
\begin{align*}
G_M(\bx)^{-1} &= I_D - \alpha^2 \frac{\nabla \ell(\bx) \nabla \ell(\bx)^\top}{1 + \alpha^2 \norm{\nabla \ell(\bx)}^2}
= I_D + (2t + t^2 \norm{\ell(\bx)}^2) \nabla \ell(\bx) \nabla \ell(\bx)^\top
\end{align*}
which is equivalent as finding the roots of the quadratic equation in $t$
\begin{align*}
\norm{\nabla \ell(\bx)}^2 t^2  + 2t+ \frac{\alpha^2}{1 + \alpha^2 \norm{\nabla \ell(\bx)}^2} = 0,
\end{align*}
whose positive root is given by
$$t_+ =   - \frac{1}{\norm{\nabla \ell(\bx)}^2} + \frac{1}{\norm{\nabla \ell(\bx)}^2\sqrt{1+ \alpha^2 \norm{\nabla\ell(\bx)}^2}}.$$ 
Setting $t = t_+$, $u = \nabla \ell(\bx)$ in $A$ and rearranging, we get
\begin{align*}
\sqrt{G_M(\bx)^{-1}} = I_D +  \frac{1}{\norm{ \nabla \ell(\bx)}^2} \bigg( \frac{1}{ L_\alpha(\bx)^\frac{1}{2}} - 1 \bigg)
\nabla \ell(\bx) \nabla \ell(\bx)^\top.
\end{align*}

Around the local modes of the log target distribution, the metric-tensor reduces to the Euclidean metric. In these cases, the scalar value $$ c(\norm{\nabla \ell(x)}^2) = \frac{1}{\norm{ \nabla \ell(\bx)}^2} \bigg( \frac{1}{ L_\alpha(\bx)^\frac{1}{2}} - 1 \bigg)$$ might cause instability in computer implementations since the norm of the gradient will be zero. 
To address this computational issue, we obtain the limit of $c(\cdot)$ when the gradient approaches the zero vector. That is, 
\begin{align*}
\lim_{\norm{\nabla \ell(x)}^2 \rightarrow 0} \frac{1}{\norm{ \nabla \ell(\bx)}^2} \bigg( \frac{1}{ (1 + \alpha^2 \norm{\nabla \ell(x)}^2)^\frac{1}{2} } - 1 \bigg) &= \lim_{\norm{\nabla \ell(x)}^2 \rightarrow 0} \frac{\frac{1}{ (1 + \alpha^2 \norm{\nabla \ell(x)}^2)^\frac{1}{2} } - 1}{\norm{ \nabla \ell(\bx)}^2} \frac{\frac{1}{ (1 + \alpha^2 \norm{\nabla \ell(x)}^2)^\frac{1}{2} } + 1}{\frac{1}{ (1 + \alpha^2 \norm{\nabla \ell(x)}^2)^\frac{1}{2} } + 1} \\
&= \lim_{\norm{\nabla \ell(x)}^2 \rightarrow 0} \frac{\frac{1}{ 1 + \alpha^2 \norm{\nabla \ell(x)}^2 } - 1}{ \norm{ \nabla \ell(\bx)}^2 \left( \frac{1}{ (1 + \alpha^2 \norm{\nabla \ell(x)}^2)^\frac{1}{2} } + 1 \right) } \\
&= \lim_{\norm{\nabla \ell(x)}^2 \rightarrow 0} - \frac{\alpha^2}{ (1 + \alpha^2 \norm{\nabla \ell(x)}^2) \left( \frac{1}{ (1 + \alpha^2 \norm{\nabla \ell(x)}^2)^\frac{1}{2} } + 1 \right)  } \\
& = - \frac{\alpha^2}{2},
\end{align*}
so that for $\norm{\nabla \ell(\bx)}^2 \approx 0$ we approximate the metric-tensor square root as $$\sqrt{G_M(\bx)^{-1}} \approx I_D - \frac{\alpha^2}{2} \nabla \ell(\bx) \nabla \ell(\bx)^\top.$$

\section{EXPERIMENT DETAILS}
\label{sec:experiments}

In this section we provide all computational details for the empirical experiments and demonstrations shown in the main paper, with some additional visualizations.

\subsection{Ring Probability Distribution (Figure 3)}

Figure~3 plotted geodesic curves of LMC-Monge for different $\alpha$. The ring distribution used here was defined as follows. Let the random variables $R \sim \mathcal{N}(\mu, \sigma^2)$ and $\Theta \sim U[0, 2\pi]$. We now define new random variables
\begin{align*}
    X &= R \cos(\Theta) \Longleftrightarrow R = \sqrt{X^2 + Y^2}\\ 
    Y &= R \sin(\Theta) \phantom{\Longleftrightarrow} \ \ \Theta = \arctan(Y/X)
\end{align*}
Since the above transformation is one-to-one and smooth, by the Jacobian method of transformation of random variables we get the distribution
\begin{align*}
    \pi_{X, Y}(x, y) &= \mathcal{N}\big(\sqrt{x^2 + y^2}|\mu, \sigma^2 \big)/(2\pi \ \sqrt{x^2 + y^2}),
\end{align*}
where the Jacobian $\big|\partial(r, \theta)/\partial(x, y)\big| = 1/\sqrt{x^2 + y^2}$. The parameter $\mu$ controls the radius of the ring measured from the origin and the parameters $\sigma^2$ the thickness of the ring. 

We used $\mu=12$ and $\sigma^2=0.12$ for Figure~3. The geodesic trajectories were integrated using $\varepsilon=0.03$ and $L_F = 200$ for one sample path, in order to guarantee smooth paths with minimal integration error. 

\subsection{Funnel (Section 5.1)}

\begin{table}[t]
\caption{Parameter settings for the funnel experiments.}
\label{tab:funnel}
\begin{center}
\begin{tabular}{l|rrrrrrr}
$D$ & 1 & 3 & 5 & 10 & 30 & 40 & 50 \\
\hline
$\varepsilon$ & 0.2 & 0.2 & 0.09 & 0.04 & 0.025 & 0.02 & 0.017\\
$L_F$ & 9 & 9 & 25 & 100 & 150 & 180 & 250
\end{tabular}
\end{center}
\end{table}

For the funnel probability distribution in Section~5.1 we used $\alpha$ $=$ $1$ for all dimensionalities $D$ and the parameters $\varepsilon$ and $L_F$ are provided in Table~\ref{tab:funnel}. These were set based on manual inspection following the basic principle of using smaller $\epsilon$ and larger $L_F$ for the more complex cases, but the results are not sensitive to the exact choices.
For HMC-Nuts we used the classical method from \citet{hoffman:14} as implemented in {\tt Turing.jl} using the command
{\tt NUTS\{SliceTS, ClassicNoUTurn\}(LeapFrog(stepsize))}, but note that other variants of the NUTS algorithm behaved in a very similar manner. Using {\tt MassMatrixAdaptor()} to fine-tune the Euclidean metric tensor $M$ made convergence faster, but did not help improving the exploration. We set the initial step size using {\tt find\underline{\phantom{a}}good\underline{\phantom{a}}stepsize()}.

\begin{figure}[t]
    \centering
    \includegraphics[width=0.99\columnwidth]{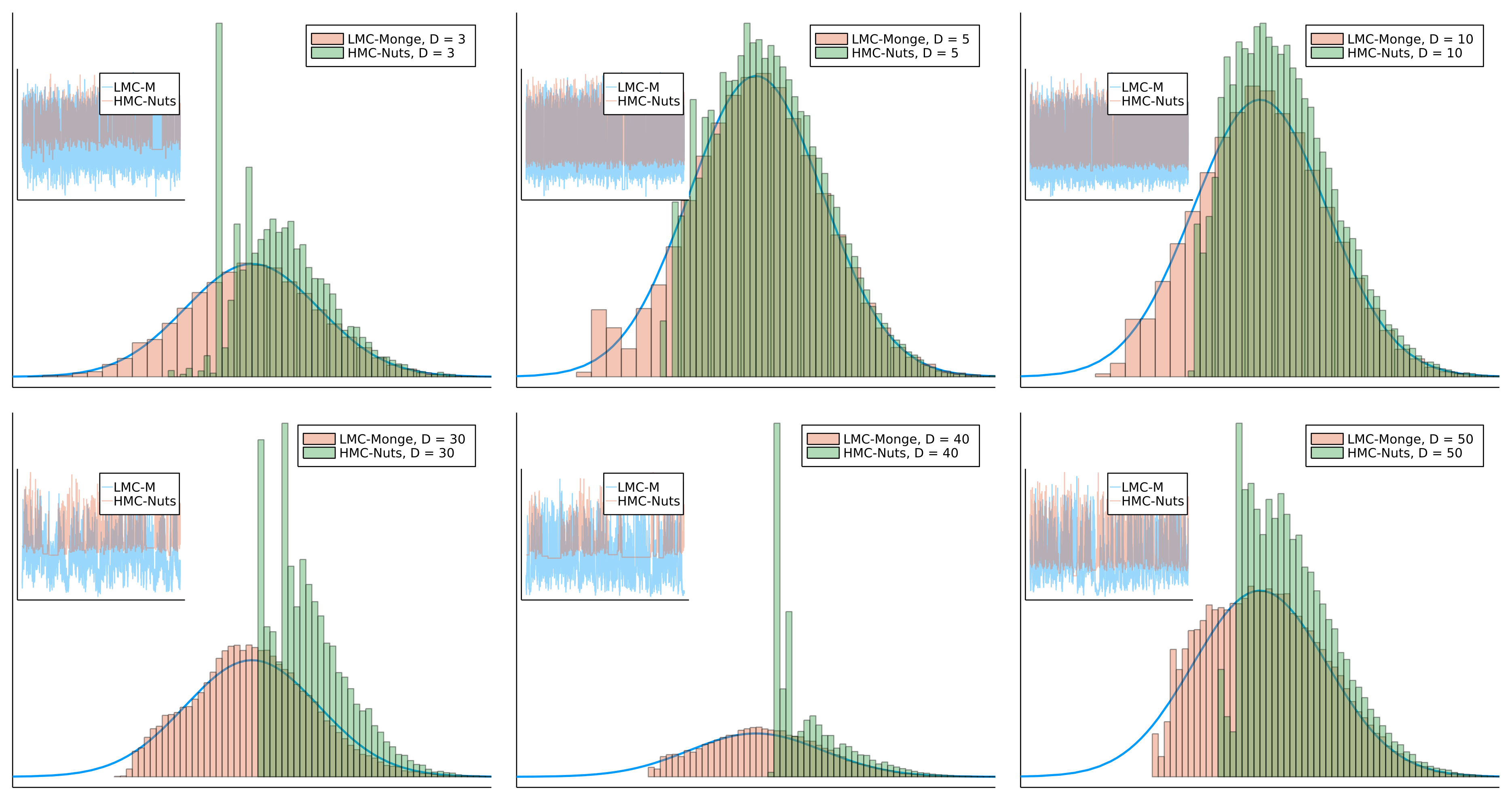}
    \caption{Histograms and MCMC chains for the LMC with Monge metrics and HMC-nuts. The geometric MCMC tends to be better in reaching corners with high curvature compared to HMC-Nuts, leading to a better estimate of the marginal distribution of the parameter $a$. The blue lines depict the true marginal distribution.}
    \label{fig:klfunall}
\end{figure}

The initial values for all cases were given by $\bx^{(1)} = 5 \mathds{1}_D$, where $\mathds{1}_D$ is the vector composed by $D$ unitary elements. Figure 5 in the main paper showed the KL divergence as function of $D$ and illustrated two margins for $D \in \{3, 5, 10, 30, 40, 50\}$. For completeness, we plot the marginals for all $D$ in Figure~\ref{fig:klfunall} to show that the difference between LMC-Monge and NUTS is consistent.

\subsection{Logistic Regression (Section 5.2)}

All the binary classification tasks in the main text use the logit link function to model the probability parameter. The un-normalised posterior distribution is given by
\begin{align*}
    \pi(\btheta|\by) \propto \prod_{i = 1}^n \left( \dfrac{\exp(\x^\top_i \btheta)}{1 + \exp(\x^\top_i \btheta)} \right)^{y_i} \left( \dfrac{1}{1 + \exp(\x^\top_i \btheta)} \right)^{n - y_i} \mathcal{N}(\btheta|\0, 100 \times I_D).
\end{align*}
where each $y_i \in \{0, 1\}$ and $\x_i$ is a vector of covariates (or inputs). The number $n$ in the sample-size and $D$ is the number of parameters in the model. This formulation, including the prior choice, matches the one used by \citet{lan:2015}.

For a sample from a Markov chain $(X_i)_{i = 1, \ldots, N}$. The effective sample size (ESS) was computed as 
\begin{align*}
    ESS = \dfrac{N}{1 + 2 \displaystyle\sum_{t = 1}^{N-2} \hat{\rho}_t}
\end{align*}
where $$\hat{\rho}_t = \dfrac{1}{N-t}\sum_{r = 1}^{N-t}(X_r - \bar{X})(X_{r + t} - \bar{X}) $$
using the implementation from the package {\tt MCMCDiagnostics.jl}.

For the LMC-Monge we chose $\varepsilon$ and $L_F$ by trial and error following the same principle that \citet{lan:2015} used for the original LMC-Fisher, aiming for acceptance probability between $0.6$ and $0.9$. For LMC-Fisher,  RMHMC-Fisher and HMC-Euclidean, we used the values provided in the Matlab implementations in {\tt https://bitbucket.org/geomstatcomp/lagrangian-monte-carlo/src/master/}, satisfying the same acceptance probability thresholds. For HMC-Nuts we again used the  {\tt find\underline{\phantom{a}}good\underline{\phantom{a}}stepsize()} function to set $\varepsilon$, but it failed to converge due to too large-step size. We fixed this by trial and error, ending up using a smaller $\varepsilon$. Note that HMC-Nuts automatically adapts the step length during the algorithm and hence the acceptance probability differs from the aimed range, and NUTS has no parameter $L_F$ as the integration length is determined by the algorithm. Table~\ref{tab:settings} lists the final values used for all methods for the experiment reported in Table~2. 
For the experiment reported in Figure 6, we used $L_F= 7$ and $\varepsilon = 0.09$ for all MCMC runs and data-sets.
\begin{table}[!t]
\caption{Parameter settings for the logistic regression experiment.}
\label{tab:settings}
\begin{center}
\begin{tabular}{llrr}
Data                 & Method       & $\varepsilon$ & $L_F$  \\ \hline\hline
Heart                & LMC-Monge ($\alpha=0.01$) & 0.085 & 7 \\
                     & LMC-Fisher   & 0.75 & 5 \\
                     & RMHMC-Fisher & 0.75 & 5 \\
                     & HMC-Euclidean   & 0.18 & 25 \\
                     & HMC-Nuts   & 0.066 & NA \\
\hline
German                & LMC-Monge ($\alpha=0.01$) & 0.05 & 5\\
                     & LMC-Fisher   & 0.8 & 5   \\
                     & RMHMC-Fisher & 0.67 & 6 \\
                     & HMC-Euclidean   & 0.063 & 64 \\
                     & HMC-Nuts   & 0.066 & NA \\
\hline
Australian                & LMC-Monge ($\alpha=0.01$) & 0.085 & 6 \\
                     & LMC-Fisher   & 0.75 & 6 \\
                     & RMHMC-Fisher & 0.75 & 6 \\
                     & HMC-Euclidean   & 0.11 & 40 \\
                     & HMC-Nuts   & 0.066 & NA \\
\hline
\end{tabular}
\end{center}
\end{table}
%
%
\subsection{Squiggle Probability Distribution (Additional experiment)}

\begin{figure}[!b]
    \centering
    \includegraphics[width = 16.5cm, height = 8.3cm]{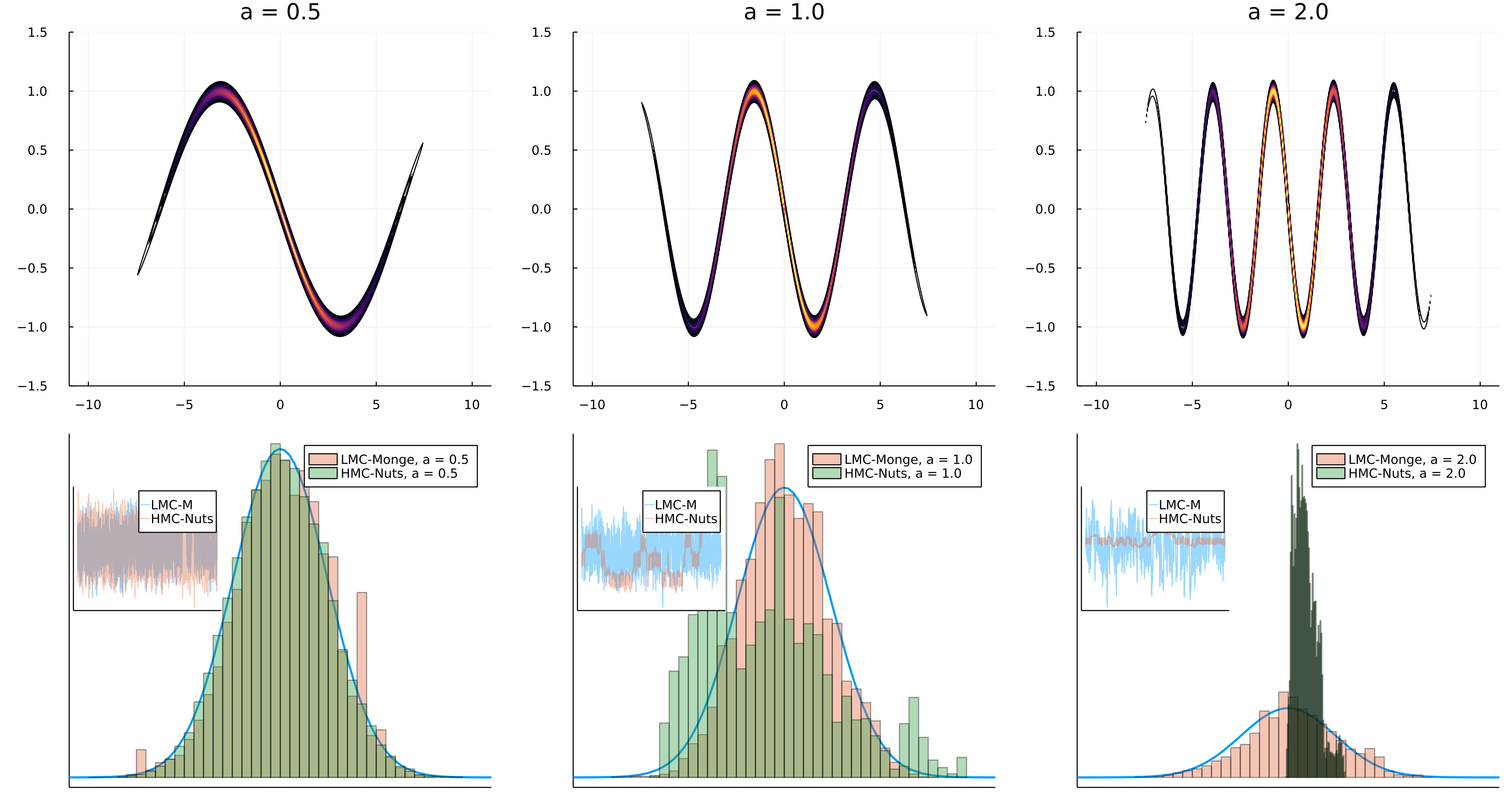}
    \caption{The first row depicts the forms of \eqref{eq:squiggle} with varying $a$. The larger the value of $a$ is, the longer is its sinusoidal form. In the second row, histograms and MCMC chains for LMC and NUTS are shown. LMC algorithm tends to be better in exploring the typical sets when compared to HMC-Nuts, leading to a better estimate of the marginal distribution $\pi_X(x_1)$, in blue.}
    \label{fig:squiggle}
\end{figure}

Here we define the squiggle probability distribution and provide extra empirical evidence for quality of the proposed algorithm based on the embedding. Let the vector of random variables $Y = (Y_1, Y_2) \sim \mathcal{N}(\0, \Sigma)$. Define the new vector $X = (X_1, X_2) = (Y_1, Y_2 - \sin(a  Y_1))$. For Jacobian method of transformation of random variables we need the inverse mapping given by $(Y_1, Y_2) = (X_1, X_2 + \sin(a X_1))$. Hence the joint distribution in $X$ reads,
\begin{align} \label{eq:squiggle}
    \pi_X(x_1, x_2|a, \Sigma) = \mathcal{N}(\by(x_1, x_2)|\0, \Sigma)|\det J_{\bx \rightarrow \by}| = \mathcal{N}(\by(x_1, x_2)|\0, \Sigma)
\end{align}
since $|\det J_{\bx \rightarrow \by}| = 1$. The parameter $a \geq 0$.
In the experiment we vary $a \in \{0.5, 1.0, 2.0 \}$, $\Sigma = [10 \ 0.01; \ 0.01 \ 0.001]$ with initial point $\bx_0 = [1, -1]^\top$. The chain size is $N = 60000$. For the LMC-monge the step-size $\varepsilon$ was $0.07, 0.07, 0.025$, the leapfrog steps $L_F$ were  $13, 13, 35$ and $\alpha = 1.0$. Those were chosen again by the inspection of the MCMC chain's convergence. For the HMC-Nuts, we set it similarly as before and we also used {\tt find\underline{\phantom{a}}good\underline{\phantom{a}}stepsize} function to set the initial $\varepsilon$ at $\bx_0$.  See Figure \ref{fig:squiggle} for the visualisation of the results.

\end{document}